\documentclass[a4paper,11pt,notoc]{article}
\usepackage{jheppub} 

\usepackage{graphicx}
\usepackage{slashed}
\usepackage{booktabs}

\title{Beyond effective field theory for dark matter searches at the LHC 
}

\preprint{IPPP/13/63, DCPT/13/128}

\author[a]{O.~Buchmueller,}
\author[b]{Matthew J.~Dolan,}
\author[b]{and Christopher McCabe}

\affiliation[a]{High Energy Physics Group, Blackett Laboratory, Imperial College, Prince Consort Road, London, SW7 2AZ, U.K.\ }
\affiliation[b]{Institute for Particle Physics Phenomenology, Durham University, South Road, Durham, DH1 3LE, U.K.\ }

 \emailAdd{oliver.buchmueller@cern.ch}
 \emailAdd{m.j.dolan@durham.ac.uk}
 \emailAdd{christopher.mccabe@durham.ac.uk}

\abstract{
We study the validity of effective field theory (EFT) interpretations of monojet searches for dark matter at the LHC for vector and axial-vector interactions.
We show that the EFT approach is valid when the mediator has mass $m_{\rm{med}}$ greater than 2.5~TeV. We find that the current limits on the contact interaction scale $\Lambda$ in the EFT apply to theories that are perturbative for dark matter mass $m_{\rm{DM}}<800$~GeV. However, for all values of $m_{\rm{DM}}$ in these theories, the mediator width is larger than the mass, so that a particle-like interpretation of the mediator is doubtful. Furthermore, consistency with the thermal relic density occurs only for $170\lesssim m_{\rm{DM}}\lesssim 520$~GeV. For lighter mediator masses, the EFT limit either under-estimates the true limit (because the process is resonantly enhanced) or over-estimates it (because the missing energy distribution is too soft). We give some `rules of thumb' that can be used to estimate the limit on $\Lambda$ (to an accuracy of $\sim 50\%$) for any $m_{\rm{DM}}$ and $m_{\rm{med}}$ from knowledge of the EFT limit. We also compare the relative sensitivities of monojet and dark matter direct detection searches finding that both dominate in different regions of the $m_{\rm{DM}}$\,--\,$m_{\rm{med}}$ plane. Comparing only the EFT limit with direct searches is misleading and can lead to incorrect conclusions about the relative sensitivity of the two search approaches.
}

\begin{document}
\maketitle
\flushbottom

\section{Introduction}
\label{sec:intro}

In many models of particle dark matter, the mechanism that generates the relic abundance requires that the dark matter interacts with the Standard Model fields. This interaction implies that dark matter may be produced from the collisions of Standard Model particles at a particle accelerator. Although the dark matter leaves the detector without depositing energy, if it recoils against a final state parton, the resultant signature is a monojet recoiling against missing energy (MET)~\cite{Cao:2009uw,Beltran:2010ww,Goodman:2010yf,Bai:2010hh,Goodman:2010ku,Fox:2011fx,Rajaraman:2011wf,Fox:2011pm,Nelson:2013pqa}.

While the monojet is a generic signature of dark matter at hadron colliders, the multitude of dark matter models makes it impossible for experiments to provide limits that are applicable to all models. It is therefore desirable to have an interpretative framework onto which a large class of models can be mapped. Such a framework should indicate how sensitive the monojet search is, while making clear its domain of validity.

So far, both the Large Hadron Collider (LHC) experiments ATLAS~\cite{ATLAS:2012ky,ATLAS:2012zim} and CMS~\cite{Chatrchyan:2012me} (updated in~\cite{CMS-PAS-EXO-12-048}) have presented their results for monojet searches using an effective field theory (EFT) framework, setting constraints on the contact interaction scale $\Lambda$. While the EFT framework is useful, providing (for example) a simple mapping of LHC limits onto the dark matter-nucleon scattering cross-section, which is constrained by direct detection experiments, care must be taken to ensure that the results within this framework are valid. This requires the particle mediating the interaction to be much heavier than the typical energy transfer. When the mediator is lighter than this, the contact interaction is resolved and effects from a UV complete theory must be taken into account. For instance, the production cross-section is enhanced if the mediator is produced on-shell and suppressed when it is off-shell, leading to an enhancement and suppression of dark matter-nucleon scattering cross-section respectively.

\sloppy
In this article we consider an alternative framework: simplified models. These have proven particularly useful in the interpretation of other Beyond the Standard Model~searches~\cite{Alves:2011wf}. Rather than considering the whole theory (the MSSM for example), constraints are set on a simple model that captures the most relevant physics being probed in that particular search. The simplified model should be chosen so that more complete UV theories can be easily mapped onto it~\cite{Alwall:2008ag,Alves:2010za,Alves:2011wf,Essig:2011qg}. In the simplified model there is no need to worry about restricting the integration over phase space and it avoids the breakdown of the EFT associated with perturbative unitarity of the contact interactions~\cite{Shoemaker:2011vi,Fox:2012ee}. While we focus on monojet searches in this paper, simplified models are more generally applicable to other di-jet, Higgs invisible width and mono-searches~\cite{Andrea:2011ws,Fox:2011fx,Bai:2012xg,Carpenter:2012rg,Lin:2013sca}, for which experimental results exist~\cite{Abdallah:2008aa,ATLAS-CONF-2013-073,CMS-PAS-EXO-13-004,CMS-PAS-HIG-13-018,ATLAS-CONF-2013-011,ATLAS-CONF-2012-148,Chatrchyan:2013qha}. These searches may provide a complementary and orthogonal set of constraints to the monojet searches that we discuss~\cite{Frandsen:2012rk}.

\fussy
Throughout, we focus on (axial)-vector mediators, comparing the search limits in the EFT with those in the simplified model. In section~\ref{sec:CMS} we discuss the CMS monojet search~\cite{Chatrchyan:2012me,CMS-PAS-EXO-12-048}, which we consider as representative for the whole class of LHC monojet searches, showing that our simulation of this search is in good agreement with their results. 

In section~\ref{sec:valid} we show that the EFT limit on $\Lambda$ only applies to theories where the mediator mass $m_{\rm{med}}$ is greater than $2.5$~TeV. Although in this case the limit applies to theories that are perturbative, the width $\Gamma$ of the mediator is very large with $\Gamma/ m_{\rm{med}} >1$, calling into question the interpretation of the mediator as particle-like excitation. Furthermore, there is only a narrow mass range $170 < m_{\rm{DM}} < 520$~GeV for which the dark matter can achieve the observed relic abundance through the thermal freeze-out mechanism with this (axial)-vector mediator. In the remainder of this section, we discuss how and why the EFT results do not apply to mediators with mass less than $2.5$~TeV

In section~\ref{sec:DD} we study the complementarity between direct detection and monojet searches. While it is typically assumed that monojet searches set a stronger limit on the spin-dependent dark matter-nucleon cross-section for $m_{\rm{DM}}\lesssim1$~TeV, this is only true for heavier mediators, whereas the direct detection searches provide stronger limits for light mediators. Therefore, we delineate the regions of parameter space where spin-dependent direct detection searches are stronger than monojet searches and {\it vice versa}. We also appeal to experimentalists to present their results in the form of limits in the plane of the mediator and dark matter masses, similar to those used for interpreting searches for supersymmetry.

We conclude in section~\ref{sec:conc}. Appendices~\ref{sec:thumb},~\ref{app:relic} and~\ref{app:scalar} contain `rules of thumb' for estimating the limit on $\Lambda$ for any value of $m_{\rm{DM}}$ and $m_{\rm{med}}$; details of our calculation for the dark matter relic density; and some issues related to simplified models for scalar mediators.

\section{Validating the CMS monojet analysis}
\label{sec:CMS}

\begin{figure}[t!]
\centering
\includegraphics[width=0.48 \columnwidth]{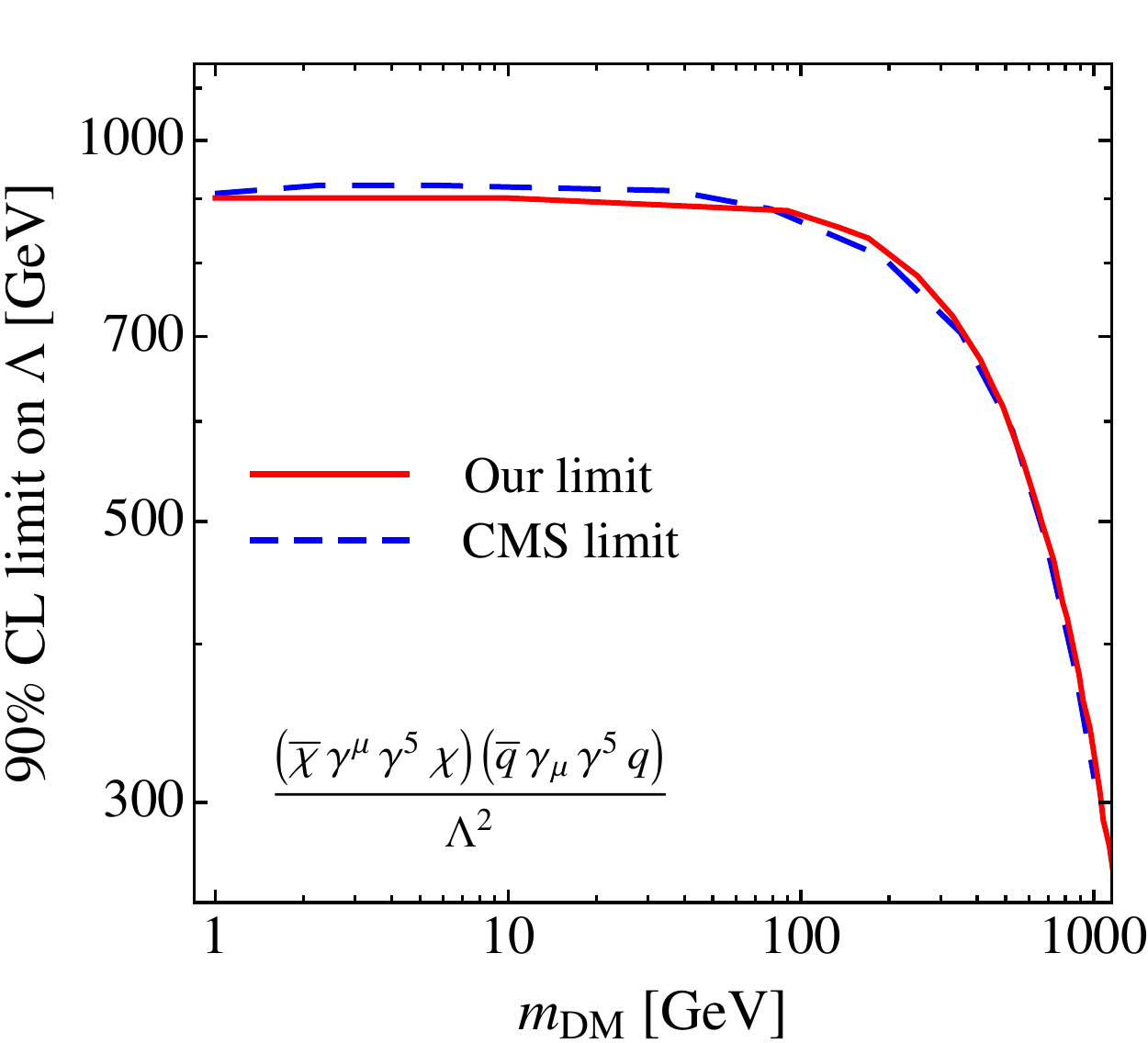}\hspace{3mm}\includegraphics[width=0.48 \columnwidth]{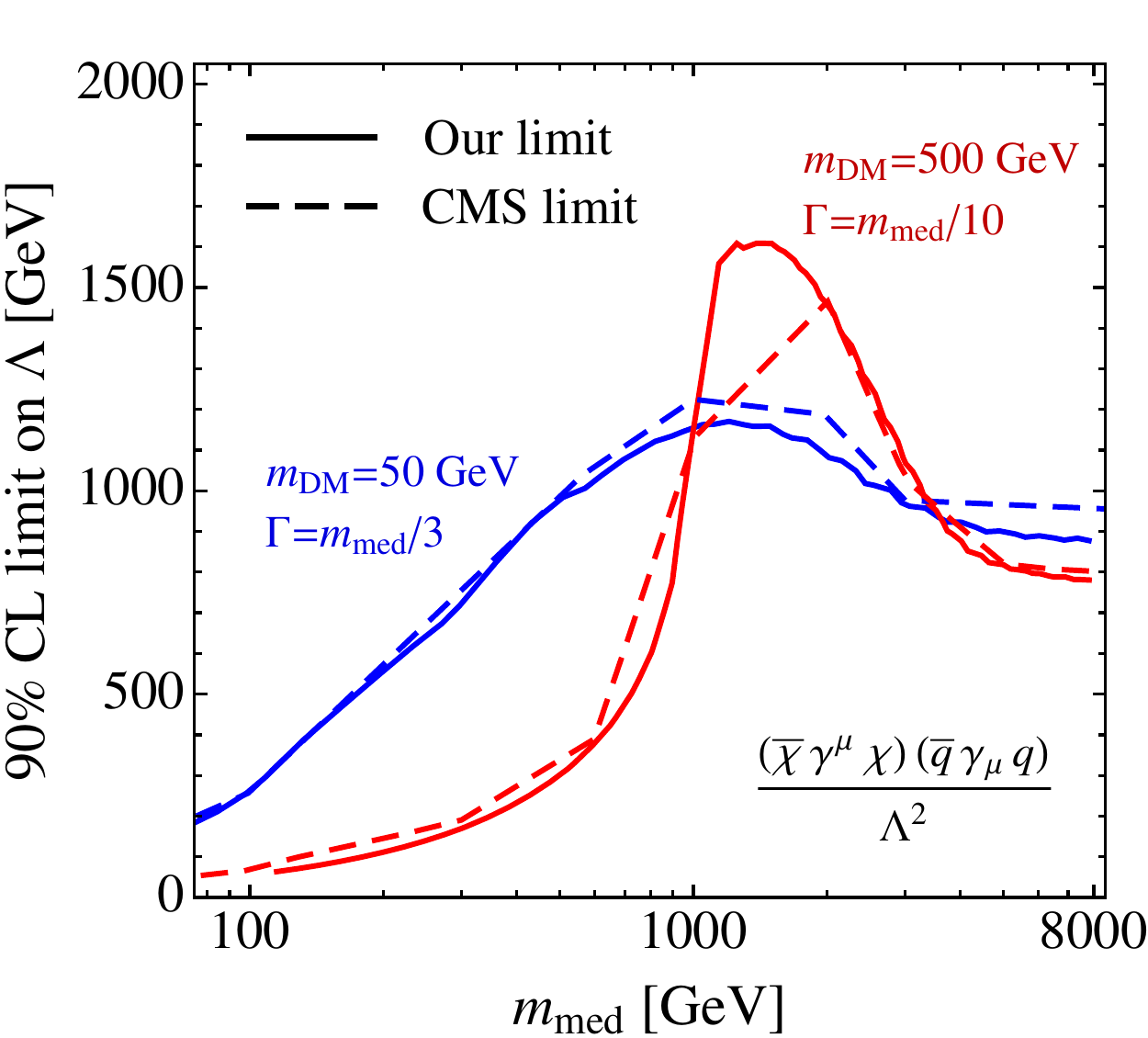}
\caption{
Left panel: A comparison of our $90\%$ CL limit (red solid) and the CMS $90\%$ CL limit (blue dashed) on the contact interaction scale $\Lambda$ as a function of $m_{\rm{DM}}$ for the axial-vector operator. The agreement is better than $5\%$. Right panel: A comparison of our $90\%$ CL limit (solid) and the CMS $90\%$ CL limit (dashed) for a vector interaction as a function of mediator mass $m_{\rm{med}}$. The blue and red lines correspond to the limit for $m_{\rm{DM}}=50$~GeV, $\Gamma= m_{\rm{med}}/3$ and $m_{\rm{DM}}=500$~GeV, $\Gamma= m_{\rm{med}}/10$ respectively, where $\Gamma$ is the mediator width. The agreement is typically better than 15\% in both cases. An exception is at the peak of the resonance, where our more fine-grained scan better resolves the peak.
}
\label{fig:CMS1}
\end{figure}

Throughout this article we make use of the CMS monojet analysis, which is based on the full data set of 19.5~fb$^{-1}$ at 8~TeV~\cite{Chatrchyan:2012me,CMS-PAS-EXO-12-048}. This search is established and well documented in the literature and it can be considered as representative for the whole class of LHC monojet searches.  In this section we describe our implementation of this search and compare our results with the CMS results. 
 
To simulate the CMS search, we use the implementation of the dark matter monojet process in MCFM 6.6~\cite{Fox:2012ru} at leading order (LO), which also includes the full effects of the mediator propagator and width.  The CMS analysis requires one hard jet with a transverse momentum $p_{\rm{T}}$ of at least 110~GeV and a total of seven signal regions with MET (or $\slashed{E}_{\rm{T}}$) greater than 250, 300, 350, 400, 450, 500 and 550~GeV were considered. The best expected limit for the dark matter search is for the MET~$>400$~GeV signal region, so only this region was used to set limits. Therefore, we restrict our implementation and validation to this particular signal region. We perform our simulations at parton level, implementing the CMS MET and geometric acceptance cuts on the jets. To account for the possibility of extra jet emission in the shower, we simulate $pp \to Z(\to \nu\bar{\nu})+1j$ using MadGraph~5~\cite{Alwall:2011uj}, showering the resulting sample using Pythia~\cite{Sjostrand:2006za} and passing it through the PGS~\cite{pgs} detector simulator with a generic LHC detector card. From this, we extract the proportion of events for which extra jets outside the CMS cuts are generated and normalise our partonic signal cross-sections with this factor. 

The results of this search were used to place limits on the contact interaction scale $\Lambda$ for a scalar, vector and axial-vector operator, assuming a Dirac fermion $\chi$ coupling with equal strength to quarks $q$. In this article, we restrict our analysis to the vector and axial-vector operators, which are
\begin{align}
\label{eq:LeffEFTvec}
\mathrm{vector:}\quad &\frac{\bar{\chi}\gamma_{\mu}\chi\,\bar{q}\gamma^{\mu}q }{\Lambda^2}\\
\text{axial-vector:}\quad&\frac{\bar{\chi}\gamma_{\mu}\gamma^5\chi \,\bar{q}\gamma^{\mu}\gamma^5 q}{\Lambda^2}
\label{eq:LeffEFT1axvec} \,.
\end{align}
Some issues related to the scalar operator are discussed in appendix~\ref{app:scalar}.

The blue dashed line in the left panel of fig.~\ref{fig:CMS1} shows the CMS 90\% CL limit on $\Lambda$ for the axial-vector operator. The red solid line shows our 90\% CL limit on this operator. Our limit and the CMS limit are in good agreement, differing by less than $5\%$ over three orders of magnitude variation in the dark matter mass $m_{\rm{DM}}$.

We have also reproduced the CMS 90\%~CL limit on $\Lambda$ for a vector interaction when the mediator is light enough to be produced on-shell at the LHC. This limit, as a function of mediator mass $m_{\rm{med}}$, is shown by the blue and red dashed lines in the right panel of fig.~\ref{fig:CMS1}. The blue and red lines show the limit for $m_{\rm{DM}}=50$~GeV, $\Gamma= m_{\rm{med}}/3$ and $m_{\rm{DM}}=500$~GeV, $\Gamma= m_{\rm{med}}/10$ respectively, where $\Gamma$ is the mediator width. The blue and red solid lines show our 90\%~CL limits. Again, we find that our limits and the CMS limits are in good agreement, differing by less than $15\%$ in both cases. The most noticeable difference is at the peak for the case $m_{\rm{DM}}=500$~GeV, $\Gamma= m_{\rm{med}}/10$ (in red). This arises because the CMS scan was less finely-grained than ours so, as expected, we better resolve the peak of the resonance.

We have thus demonstrated that our implementation is fully sufficient to reproduce the CMS analysis in the case where EFT holds (left panel of fig.~\ref{fig:CMS1}) and in the case where the mediator is light enough to be accessible at the LHC, for a variety of dark matter masses and mediator widths (right panel of fig.~\ref{fig:CMS1}).

\section{Effective field theory and beyond}
\label{sec:valid}

So far, monojet searches have typically been interpreted in the EFT framework, which is particularly simple because details of the particle mediating the interaction do not have to be specified. In this section, we will quantify when the (axial)-vector limit on the scale $\Lambda$ in the EFT framework is applicable, an area that has received relatively little attention, as well as quantifying where and when this framework breaks down.

\begin{figure}[t!]
\centering
\hspace{8mm}\includegraphics[width=0.435\columnwidth]{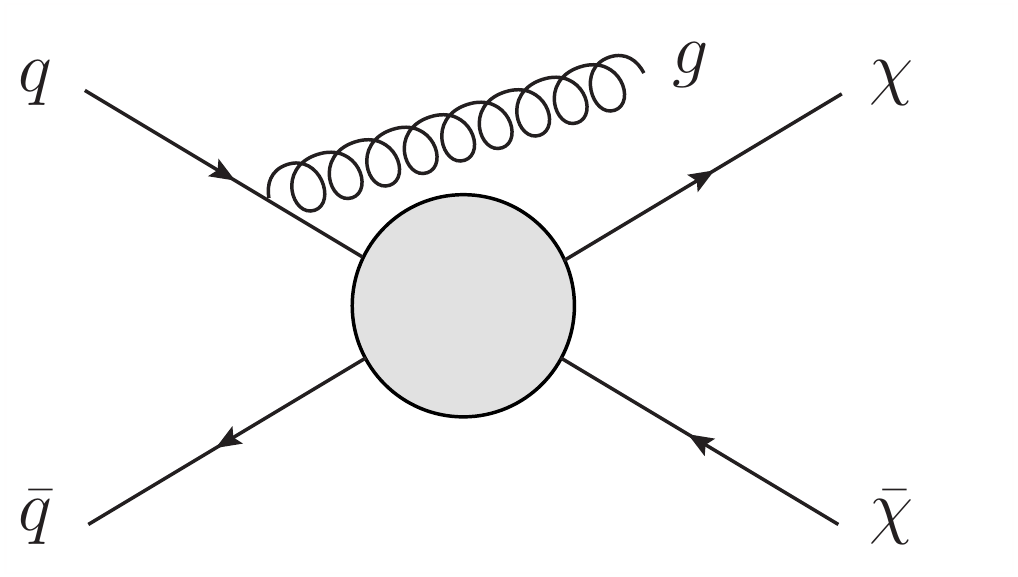}\hspace{5mm}\includegraphics[width=0.464\columnwidth]{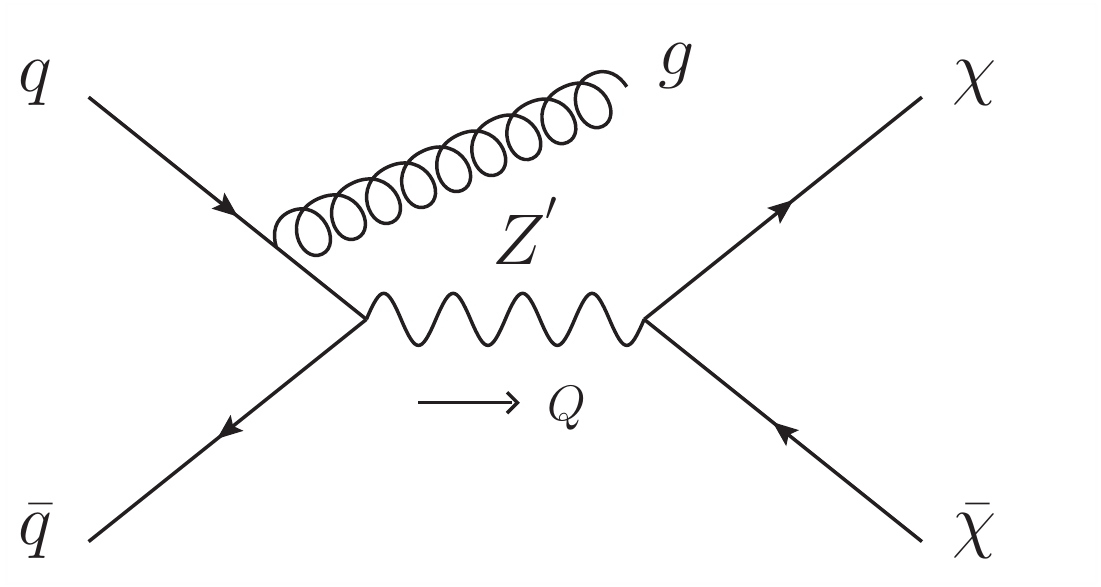}
\caption{Left panel: The monojet process from a $q\bar{q}$ initial state in the EFT framework. The contact interaction is represented by the shaded blob. Details of the particle mediating the interaction do not have to be specified. Right panel: This shows a UV resolution of the contact interaction for an (axial)-vector mediator $Z^{'}$, exchanged in the $s$-channel. The momentum transfer through the $s$-channel is denoted by $Q$. }
\label{fig:SMS}
\end{figure}

In order to go beyond the EFT framework, we must resolve the contact interaction, indicated by the shaded blob in the left panel of fig.~\ref{fig:SMS}. Upon resolving this blob, we are immediately faced with two choices: the mediator could be exchanged in the $t$-channel, in which case it must be coloured, or in the $s$-channel, in which case it may be colour neutral. In this article, we assume that it is exchanged in the $s$-channel, as in the right panel of fig.~\ref{fig:SMS}. The phenomenology of $t$-channel mediators has been explored in~\cite{Chang:2013oia,An:2013xka,Bai:2013iqa,DiFranzo:2013vra}.

While there is no unique UV resolution of the contact interaction, the simplified model that we propose contains all of the elements that we expect to appear for a mediator that is exchanged in the $s$-channel. We remain agnostic to the precise origin of the vector mediator and its coupling with dark matter and quarks. One example of such a mediator is a (axial)-vector $Z^{'}$, a massive spin-one vector boson from a broken $U(1)^{'}$ gauge symmetry~\cite{Holdom:1985ag,Babu:1997st}. A second example is a composite vector mediator, similar to the $\omega$ in QCD~\cite{Foadi:2008qv}. In either case, in addition to the usual terms in the Standard Model Lagrangian, the Lagrangian with general quark interaction terms is
\begin{equation}
\begin{split}
\mathcal{L}&=-\frac{1}{4}Z'_{\mu\nu}Z^{'\mu\nu}+\frac{1}{2}m_{\rm{med}}^2 Z^{' \mu}Z^{'}_{\mu}  + i\bar{\chi}\gamma^{\mu}\partial_{\mu}\chi-m_{\rm{DM}}\bar{\chi}\chi\\
&+Z^{'}_{\mu} \bar{\chi}\gamma^{\mu}(g_{\chi V} -g_{\chi A} \gamma^5)\chi +Z^{'}_{\mu}\sum_q\bar{q}\gamma^{\mu}(g_{q V} -g_{q A} \gamma^5)q\;.
\label{eq:pseudovectorSMS}
\end{split}
\end{equation}
Here $m_{\rm{med}}$ is the (axial)-vector mass term and $g_V$ and $g_A$ are the vector and axial couplings respectively. The dark matter particle $\chi$ is a Dirac fermion with mass $m_{\rm{DM}}$, neutral under the Standard Model gauge groups. The sum extends over all quarks and for simplicity, we assume that the couplings $g_{q V}$ and $g_{q A}$ are the same for all quarks. While in general, a $Z^{'}$ from a broken $U(1)^{'}$ will also have couplings to leptons and gauge bosons, we do not consider them here as they are not relevant for the monojet search.\footnote{We assume that the charges are chosen so the $U(1)^{'}$ gauge symmetry is anomaly free. This may require additional particles.} This simplified model is similar (albeit simpler) to the model discussed in~\cite{Frandsen:2012rk}. Simplified models of vector mediators have also been discussed in~\cite{Bai:2010hh,Shoemaker:2011vi,Frandsen:2012rk,An:2012ue,An:2012va}.

While the above Lagrangian allows for both vector and axial-vector interactions, the phenomenology and limits from the monojet search are similar in both cases. Therefore for the purposes of clarity, we focus on one: the axial-vector interaction. In the remainder of this article, we set $g_{\chi V}=g_{q V}=0$ and redefine $g_{\chi}\equiv g_{\chi A}$ and $g_{q}\equiv g_{g A}$. The axial-vector interaction has two advantages. Firstly, this interaction is non-zero for Majorana dark matter (the normalisation of our results would change by a factor of four in this case), unlike the vector interaction, which vanishes for Majorana dark matter. Secondly, the comparison between the monojet limits and direct detection searches is more interesting in this case (we consider this further in section~\ref{sec:DD}).

If the axial-vector mediator is suitably heavy (to be quantified more carefully below) it can be integrated out to obtain the effective axial-vector contact operator in eq.~\eqref{eq:LeffEFT1axvec}. In this case, the contact interaction scale is related to the parameters entering the Lagrangian eq.~\eqref{eq:pseudovectorSMS} by
\begin{equation}
\Lambda\equiv \frac{m_{\rm{med}}}{\sqrt{g_q \,g_{\chi}}}\;.
\end{equation}
In fact, even when we study the effects beyond the EFT framework, we will still use this as our definition of $\Lambda$.

\begin{figure}[t]
\centering
\includegraphics[width=0.47 \columnwidth ]{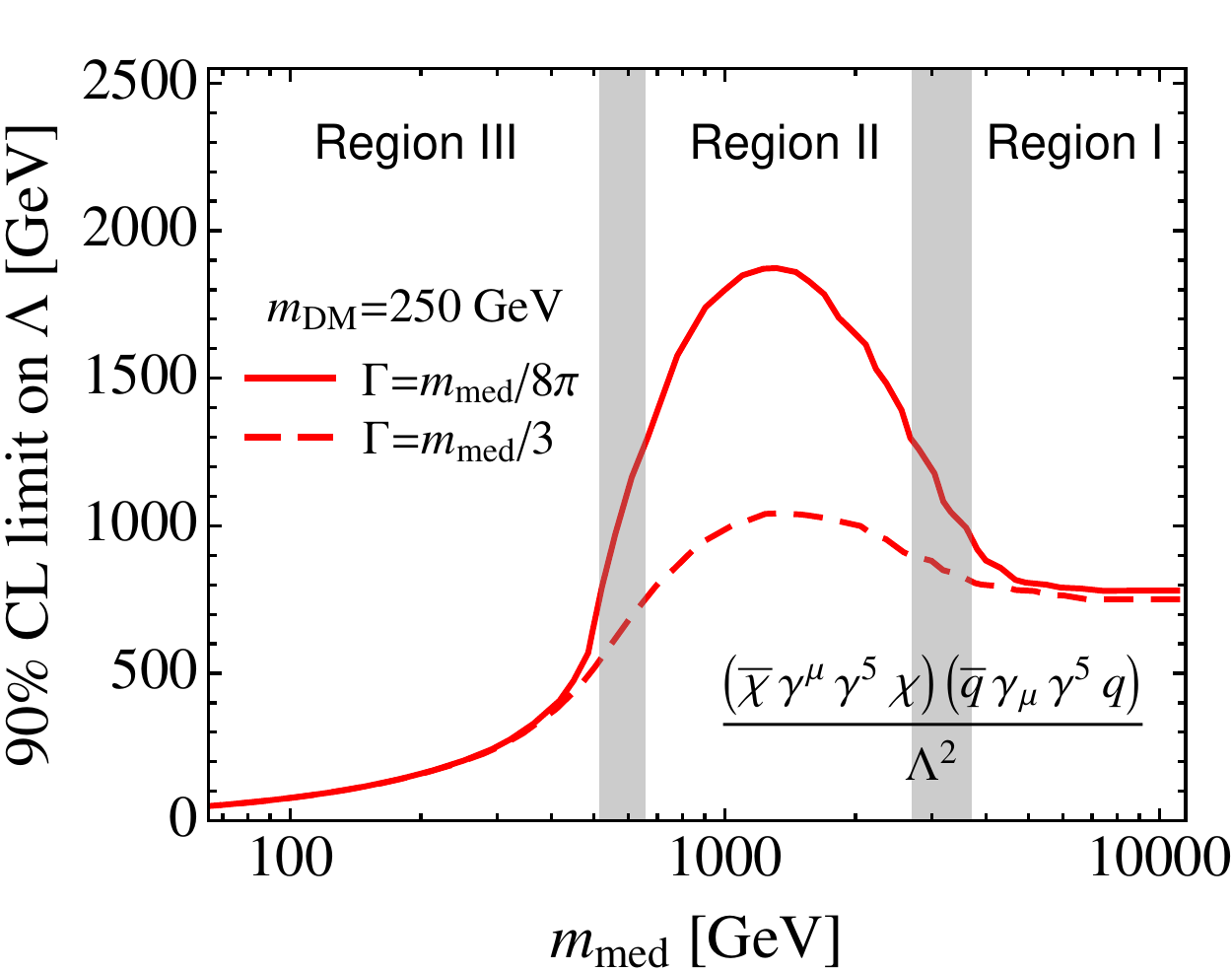}\hspace{1mm}
\includegraphics[width=0.5 \columnwidth] {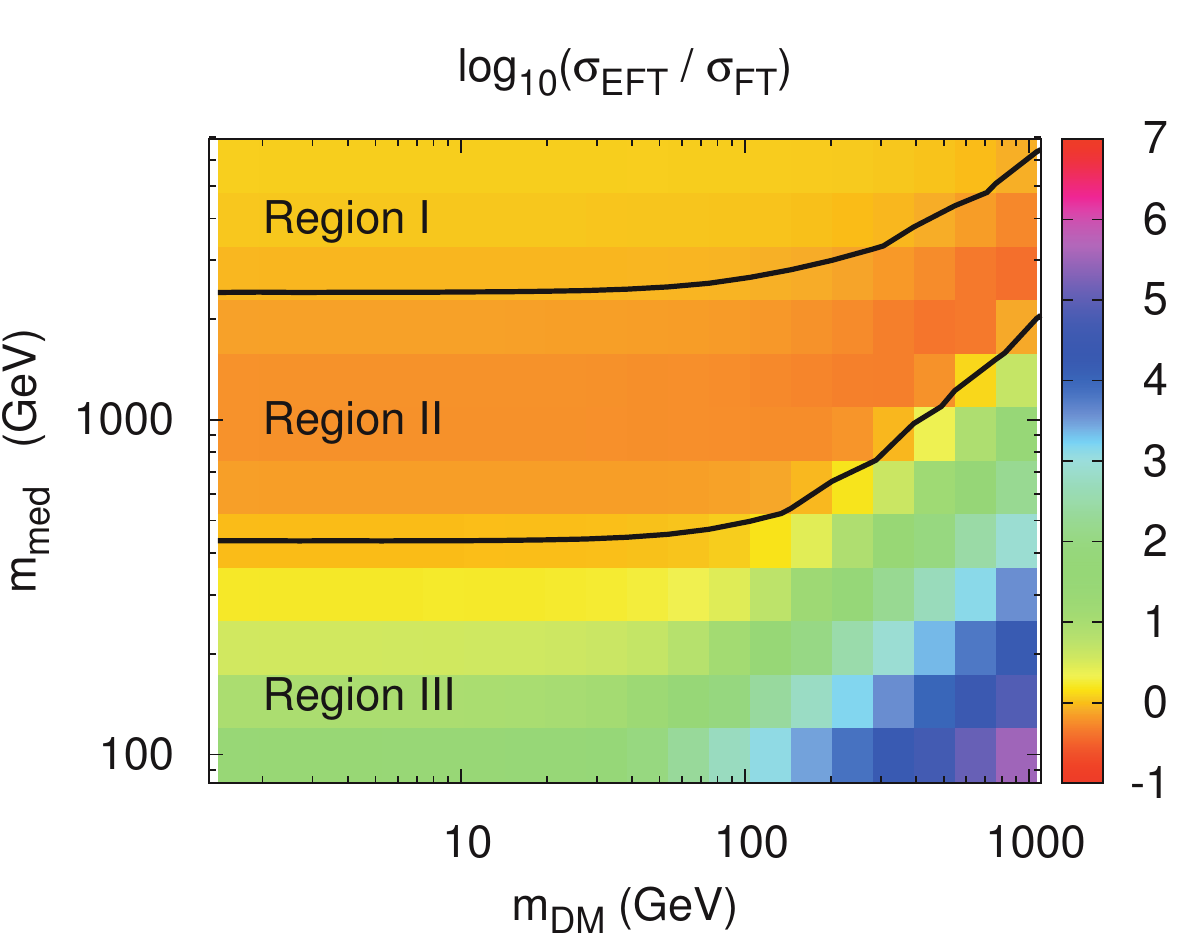}
\caption{Left panel: The 90\% CL limit on $\Lambda$ as a function of $m_{\rm{med}}$ for our axial-vector simplified model with $m_{\rm{DM}}=250$~GeV. Right panel: The ratio of the inclusive cross-sections in the EFT to the simplified model. In both panels, three distinct regions of parameter space are marked: In Region I, the EFT and simplified model calculation agree at the level of 20\% or better; in Region II, the simplified model cross-section is larger than the EFT cross-section owing to a resonant enhancement; and in Region III, the simplified model cross-section is smaller than the EFT cross-section. In the left panel we consider two mediator widths $\Gamma$. The grey shaded regions indicate that the boundary between the regions is weakly dependent on $\Gamma$.}
\label{fig:ratio}
\end{figure}

Now that we have completed the definition of the simplified model, we examine the differences between the EFT and simplified model. We first consider the specific case with $m_{\rm{DM}}=250$~GeV in the left panel of fig.~\ref{fig:ratio}, which shows the limit on $\Lambda$ as a function of $m_{\rm{med}}$. Three distinct regions of parameter space can clearly be seen: we define Region~I to be the region where the EFT and simplified model limits on $\Lambda$ agree at the level of 20\% or better (this region was studied in~\cite{Busoni:2013lha} for the scalar interaction). The measure of 20\% corresponds to the uncertainty on the signal cross-sections in CMS monojet analysis and it is used by us to determine the validity of the EFT approach~\cite{CMS-PAS-EXO-12-048}. This is the region where the EFT limit on $\Lambda$ can be applied to the simplified model and requires $m_{\rm{med}}\gtrsim3$~TeV. In Region~II, the limit on $\Lambda$ in the simplified model is larger than the EFT limit owing to a resonant enhancement. Finally, we define Region~III to be the region where the limit on $\Lambda$ in the simplified model is smaller than the EFT limit. 

We have also considered two different widths for the mediator. The width of an axial-vector mediator decaying to Dirac fermions $f$ and $\bar{f}$ with coupling $g_f$ is
\begin{equation}
\frac{\Gamma}{m_{\rm{med}}}=\frac{N_C\, g_{f}^2}{12 \pi}  \left(1- \frac{4 m_f^2}{m_{\rm{med}}^2} \right)^{3/2}\;,
\label{eq:width}
\end{equation}
where $N_C=3$ for coloured particles and is $1$ otherwise. The solid red line shows the result for a narrow width, $\Gamma=m_{\rm{med}}/8\pi$, and the dashed line for a broad width, $\Gamma=m_{\rm{med}}/3$. In Regions I and III the limit on $\Lambda$ is only weakly dependent on the width, since in both these regions, the mediator is being produced off-shell. Conversely, in Region II, the limit is strongly dependent on the width as the production is resonantly enhanced. Finally, the grey regions show that the value of $m_{\rm{med}}$ at the transitions between the different regions may change by $\sim10\%$, depending on the width.

We now consider the more general case. In the right panel of fig.~\ref{fig:ratio} we show the ratio of the inclusive cross-section (i.e.\ we take the minimum cut used by CMS, $p_{\mathrm{T},j}>110$~GeV) in the EFT, $\sigma_{\rm{EFT}}$, to that in the simplified model (or full theory, FT), $\sigma_{\rm{FT}}$, as a function of $m_{\rm{DM}}$ and $m_{\rm{med}}$. For simplicity, we have set $g_{\chi}=g_{q}=1$ so that $\Lambda = m_{\rm{med}}$ and we have calculated the width for each value of $m_{\rm{DM}}$ and $m_{\rm{med}}$ using eq.~\eqref{eq:width}. For different couplings, the width will be different and the boundaries between the regions may change by $\sim10$\% but otherwise, the plot will be similar. The orange and red regions indicate when the EFT cross-section is smaller than in the simplified model, while the green and bluer colours indicate the inverse. The same three distinct regions of parameter space can again be seen. For $m_{\rm{DM}}\lesssim 100$~GeV, we require $m_{\rm{med}}>2.5$~TeV to be in Region~I, where the EFT limit on $\Lambda$ can be used. For larger values of $m_{\rm{DM}}$, the value of $m_{\rm{med}}$ at the boundary between Region~I and~II increases, reaching $m_{\rm{med}}=6$~TeV for $m_{\rm{DM}}=1$~TeV.

We now discuss each of these regions in further detail.

\subsection{Region I: Very heavy mediator - EFT limit applies}
\label{sec:heavymed}

In Region~I, the cross-section in the simplified model and EFT agree within experimental uncertainties ($20\%$) and the limit on $\Lambda$ is independent of $m_{\rm{med}}$. This behaviour can be simply understood: expanding the propagator (while ignoring the width) for the $s$-channel resonance in powers of $Q^2/m_{\rm{med}}^2$, where $Q^2$ is the momentum transfer through the $s$-channel (see right panel of fig.~\ref{fig:SMS}), we obtain
\begin{equation}
\frac{g_q\, g_{\chi}}{Q^2-m_{\rm{med}}^2}\approx-\frac{g_q\, g_{\chi}}{m_{\rm{med}}^2}\left( 1+\frac{Q^2}{m_{\rm{med}}^2}+\mathcal{O}\left(\frac{Q^4}{m_{\rm{med}}^4}\right)\right)\;.
\label{eq:prop1}
\end{equation}
We recognise the first term outside the brackets as the contact interaction scale of the EFT: $1/\Lambda^2=g_q\, g_{\chi}/m_{\rm{med}}^2$. The EFT is valid so long as the effects of the rest of the expansion beyond leading order are small, i.e.\ if $ m_{\rm{med}}\gg Q$. At the 8~TeV LHC run, $\langle Q^2\rangle^{1/2}$ is always larger than 500~GeV~\cite{Busoni:2013lha}, so we expect $m_{\rm{med}}$ to be TeV scale in order that $m_{\rm{med}}\gg Q$. This is confirmed by the right panel of fig.~\ref{fig:ratio}, where we see that $m_{\rm{med}}$ should be at least 2.5~TeV in order that $\sigma_{\rm{EFT}}$ and $\sigma_{\rm{FT}}$ agree to better than 20\%.

Stating the minimum mediator mass $m_{\rm{med}}$ needed for the EFT limit to be valid, rather than a minimum value of $\Lambda$, is much more natural in the simplified model framework. While we can define a unique mass $m_{\rm{med}}$ for which the EFT is valid, there is not a unique scale $\Lambda$ corresponding to this mass, since there are many points in $\Lambda$--$g_q g_{\chi}$ space which map onto the same value of $m_{\rm{med}}$.

\begin{figure}[t]
\centering
\includegraphics[width=0.6 \columnwidth]{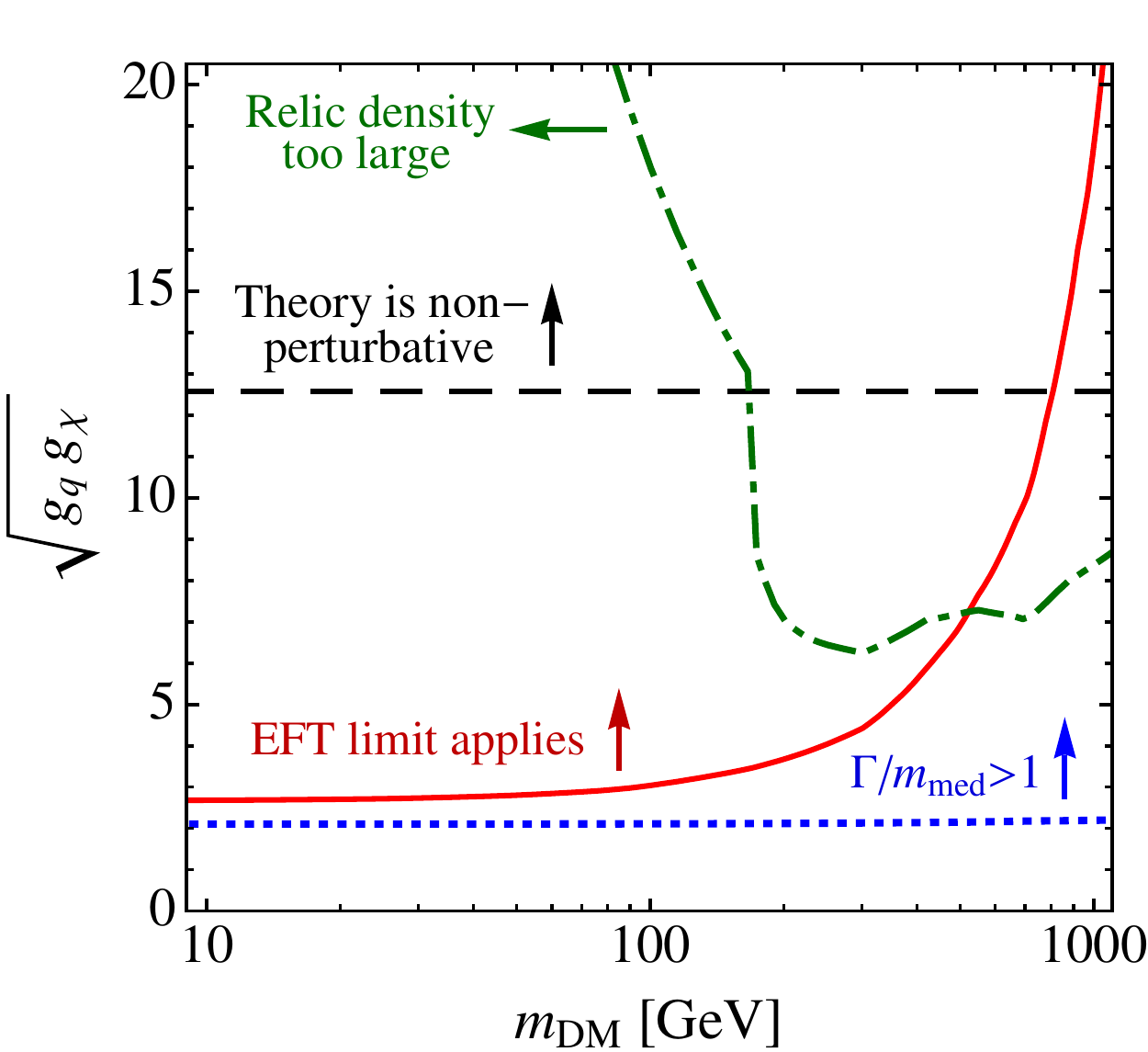}
\caption{The solid red line indicates the minimum coupling $\sqrt{g_q\,g_{\chi}}$ in order that the CMS EFT limit on $\Lambda$ applies to the simplified model. The perturbative limit on the couplings (4$\pi$) is indicated by the dashed black line. The EFT limits apply to perturbative theories for $m_{\rm{DM}}<800$~GeV. The mediator width $\Gamma$ equals its mass $m_{\rm{med}}$ when $\sqrt{g_q\,g_{\chi}}$ takes the values indicated by the dotted blue line. The EFT limits only apply to theories where $\Gamma>m_{\rm{med}}$, so the mediator may not be identified as a particle. The green dot-dashed line indicates the coupling $\sqrt{g_q\,g_{\chi}}$ where the relic density matches the observed value. This occurs in the range $170\lesssim m_{\rm{DM}}\lesssim520$~GeV.}
\label{fig:couplings}
\end{figure}

Having established the minimum mediator mass required for the EFT limit to be valid, we now elucidate the theories that are excluded by the EFT limit on $\Lambda$. First, we calculate the minimum coupling $\sqrt{g_q \, g_{\chi}}=m_{\rm{med}}/\Lambda$ that the simplified model must have for the EFT limits to apply. This is shown by the solid red line in fig.~\ref{fig:couplings}. To calculate this line, we use the CMS upper limit on $\Lambda$ from the left panel of fig.~\ref{fig:CMS1}, and the upper contour delineating the boundary between Region~I and Region~II in the right panel of fig.~\ref{fig:ratio}, giving us the minimum value of $m_{\rm{med}}$. We now make a number of comments about this region. 

The first observation is that the EFT limit rules out theories with large couplings $\sqrt{g_q\,g_{\chi}}\gtrsim3$. At larger $m_{\rm{DM}}$, this coupling is even larger because the limit on $\Lambda$ decreases while $m_{\rm{med}}$ increases. Theories are normally said to be perturbative so long as the product of the couplings $\sqrt{g_q \, g_{\chi}}$ is smaller than $4\pi$, which we have indicated by the black dashed line in fig.~\ref{fig:couplings}. From fig.~\ref{fig:couplings}, we see that theories for which the EFT limits apply are perturbative so long as $m_{\rm{DM}}<800$~GeV. 

Secondly, we find that everywhere, the mediator width is larger than the mass. For constant $g_{q}\, g_{\chi}$, the minimum width is
\begin{equation}
\frac{\Gamma_{\rm{min}}}{m_{\rm{med}}}=  g_{q\,} g_{\chi} \frac{\sqrt{3}}{6 \pi} \sqrt{\left( 1-\frac{4 m_{\rm{DM}}^2}{m_{\rm{med}}^2}\right)^{3/2}\sum_{q}\left( 1-\frac{4 m_{q}^2}{m_{\rm{med}}^2}\right)^{3/2}}\,.
\end{equation}
The dotted blue line in fig.~\ref{fig:couplings} indicates the values of $\sqrt{g_{q\,} g_{\chi}}$ for which $\Gamma_{\min}=m_{\rm{med}}$. Thus, although the EFT limit applies to theories that are perturbative, the very large mediator width means that it is unlikely that the mediator would be identified as a particle and it would be difficult to find in di-jet searches, which typically search for a narrow resonance.

The final comment concerns the dark matter relic abundance. The green dot dashed line in fig.~\ref{fig:couplings} indicates the couplings for which the relic abundance saturates the observed value, for which we take the best fit value $\Omega_{\chi\bar{\chi}} h^2=0.119$ obtained by Planck (in the minimum $\Lambda$CDM model)~\cite{Ade:2013zuv}. It is only in the small range $170\lesssim m_{\rm{DM}}\lesssim520$~GeV that this occurs (for perturbative theories). However, this line should be taken as indicative, since it assumes that only the axial-vector operator is operative and ignores effects such as co-annihilation. Over much of the parameter space, the relic abundance is too large. Therefore, additional annihilation channels or another lighter mediator is required to reduce the abundance to an acceptable level~\cite{Boehm:2003hm}. Details of this calculation can be found in appendix~\ref{app:relic}. 

Therefore, we conclude that the CMS limit on $\Lambda$ rules out rather baroque theories of dark matter. Although the theory is perturbative, the couplings required are so large that in the entire parameter space, the width of the mediator is larger than its mass $\Gamma/m_{\rm{med}}>1$. Finally, it is only for masses between 170 and 520~GeV where the dark matter can be a thermal relic (assuming there are no other interactions or particles beyond the simplified model).

\subsection{Region II: Resonant enhancement - EFT limit conservative}
\label{sec:intermed}
\begin{figure}[t]
\centering
\includegraphics[width=0.5 \columnwidth]{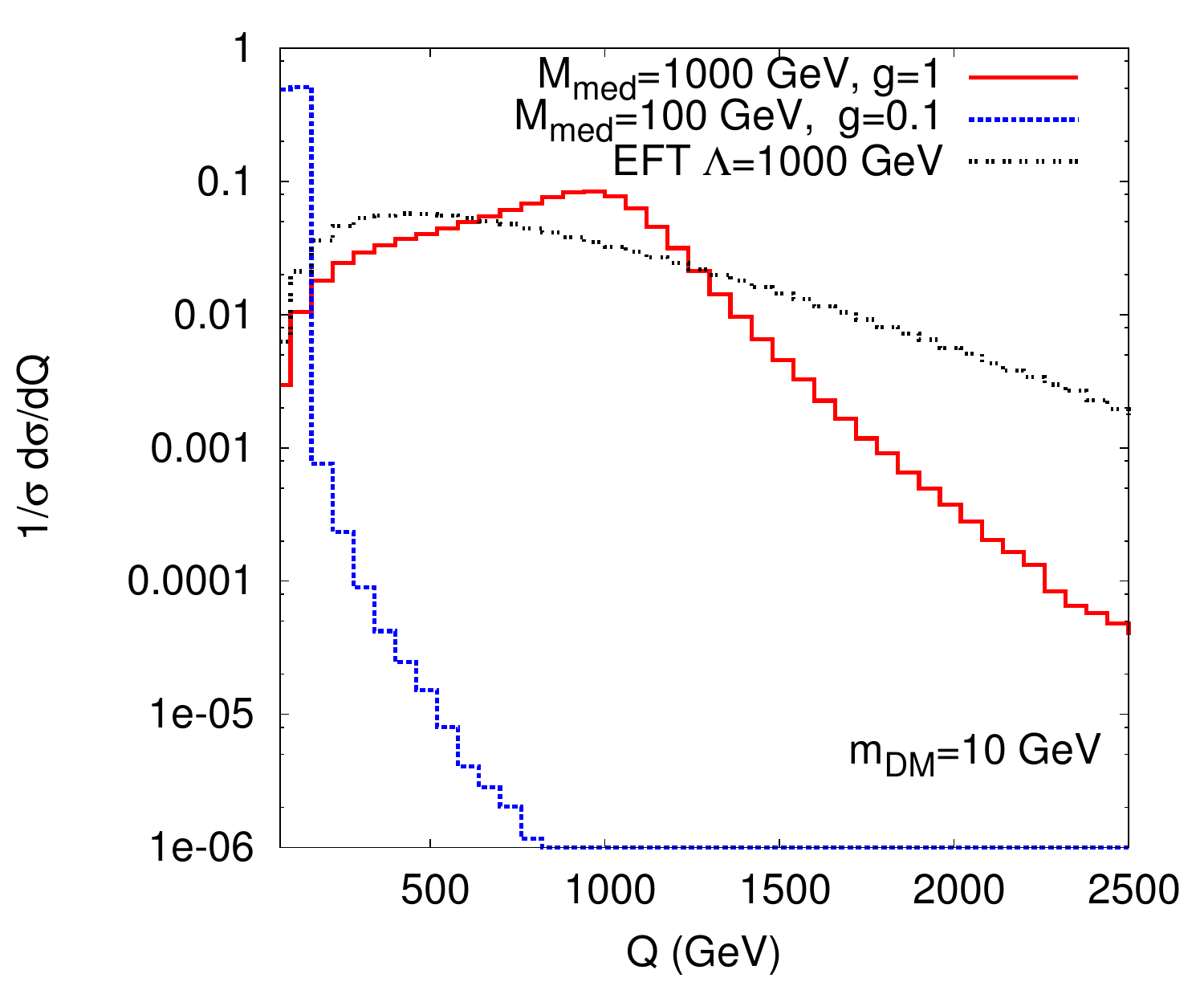}\includegraphics[width=0.5 \columnwidth]{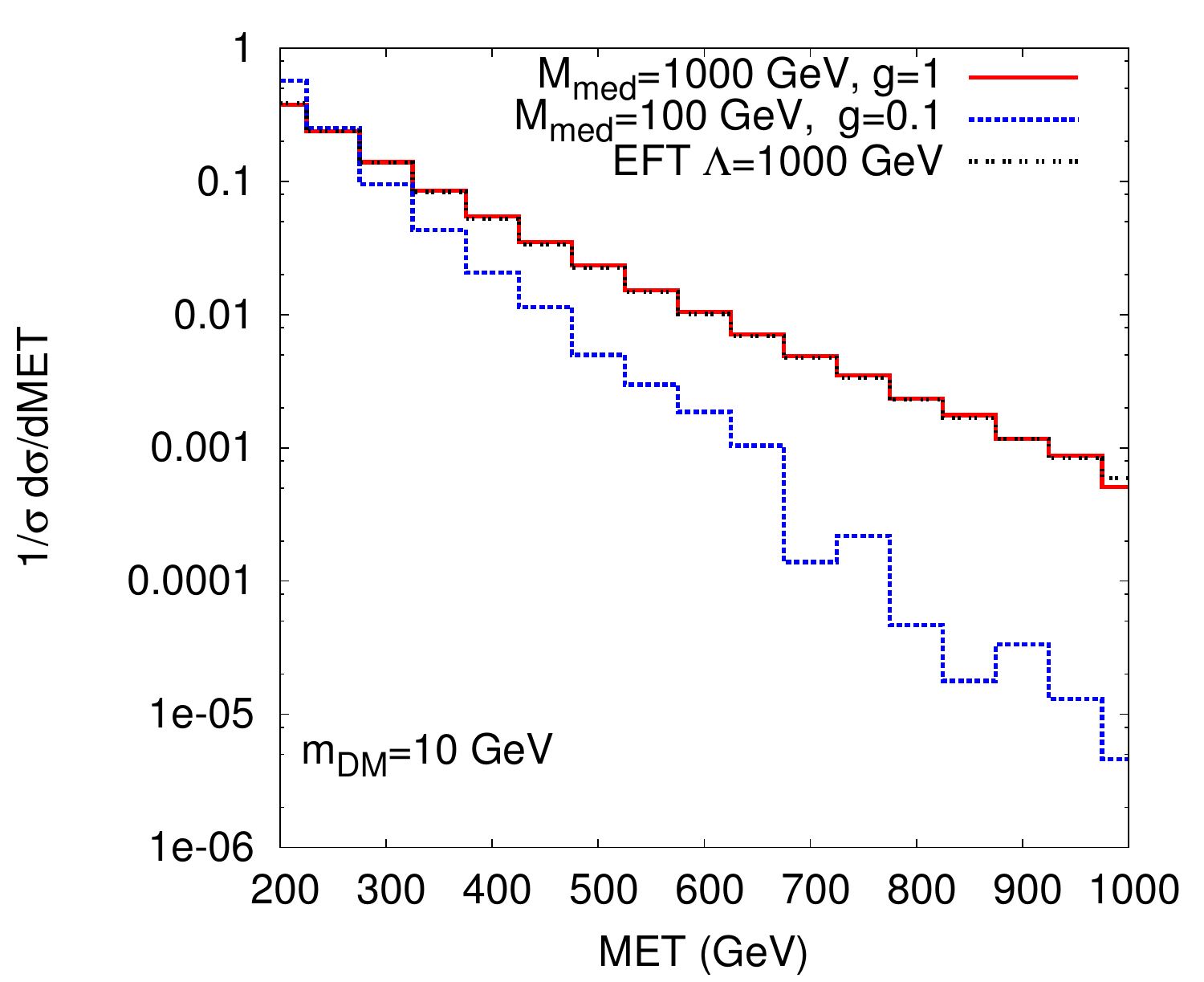}
\caption{
Left panel: The differential cross-section with respect to $Q$, the momentum transfer through the $s$-channel, or equivalently, the (unobservable) invariant mass of the dark matter pair. The dotted blue and solid red lines are for $m_{\rm{med}}=100$~GeV and 1000~GeV respectively and the resonant peak at $Q=m_{\rm{med}}$ is clearly observable. The double dotted black line shows the EFT cross-section with $\Lambda=1000$~GeV. No resonant peak is observed in the EFT limit and the cross-section extends to much larger values of $Q$. Right panel: The missing energy distribution for the same mediator masses and couplings as the left panel. The MET distribution for $m_{\rm{med}}=100$~GeV is much softer so that fewer events pass the CMS $\rm{MET}>400$~GeV cut, leading to a weaker limit on $\Lambda$. In both panels, we have fixed $g=g_q=g_{\chi}$ and $m_{\rm{DM}}=10$~GeV.}
\label{fig:PV1}
\end{figure}

In Region~II, the limit on $\Lambda$ is always larger than the EFT limit, i.e.\ the EFT limit is always conservative. This is because the cross-section is resonantly enhanced when the mediator is on-shell. This resonant enhancement can clearly be seen in the red and blue lines in the left panel of fig.~\ref{fig:PV1}, which shows a clear peak in the differential cross-section (with respect to $Q$) when $Q\approx m_{\rm{med}}$. In contrast, no resonance peak is observed in the EFT limit.

The mediator can be produced on-shell when the relation $m_{\rm{med}}^2\gtrsim4 m_{\rm{DM}}^2+\slashed{E}^2_{\rm{T}}$ is satisfied~\cite{Fox:2012ru}. Taking $\slashed{E}_{\rm{T}}=400$~GeV, the cut imposed by CMS in their search, we find that this relationship gives a useful `rule of thumb' to determine the boundary between Region~II and~III in fig.~\ref{fig:ratio}. The accuracy of this rule and other similar rules are discussed in appendix~\ref{sec:thumb}.

The left panel of fig.~\ref{fig:ratio} demonstrates that the mediator width has a large impact on the limit on $\Lambda$. At the peak of the resonance, the limit on $\Lambda$ increases by a factor of 1.8 between $\Gamma=m_{\rm{med}}/3$ and $\Gamma=m_{\rm{med}}/8\pi$.  Another `rule of thumb' is that the limit on $\Lambda$ at the peak scales as $\Gamma^{-1/4}$ (this scaling is exact for an $e^+ e^-$ collider~\cite{Fox:2011fx}). In this case, this rule of thumb predicts that at the peak, the limit on $\Lambda$ would differ by 1.7.

\subsection{Region III: Light mediator - EFT limit too strong}
\label{sec:lightmed}

In Region~III, the limit on $\Lambda$ is smaller than the EFT limit. We observe from the left panel of fig.~\ref{fig:ratio} (and right panel of fig.~\ref{fig:CMS1}) that the limit on $\Lambda$ in this region is approximately proportional to $m_{\rm{med}}$. This occurs because the limit on $g_q \,g_{\chi}$ is (approximately) constant with respect to $m_{\rm{med}}$ in this region, so that $\Lambda\, \left(=m_{\rm{med}}/\sqrt{g_q\,g_{\chi}}\right)$ depends linearly on $m_{\rm{med}}$. We can understand this behaviour by examining the propagator in the limit $Q^2\gg m_{\rm{med}}^2$
\begin{equation}
\frac{g_q\,g_{\chi}}{Q^2-m_{\rm{med}}^2} \approx \frac{g_q\,g_{\chi}}{Q^2} \left(1 + \frac{m_{\rm{med}}^2}{Q^2} +\mathcal{O}\left( \frac{m_{\rm{med}}^4}{Q^4}\right)  \right) \; .
\end{equation}
Since $\langle Q^2 \rangle^{1/2}$ is always greater than 500~GeV~\cite{Busoni:2013lha} (this can also be seen from the black dotted line in the left panel of fig.~\ref{fig:PV1}), we find that the higher order terms involving $m_{\rm{med}}$ are suppressed for $m_{\rm{med}}\ll 500$~GeV. This implies that the limit on $g_q \,g_{\chi}$ is approximately constant with respect to $m_{\rm{med}}$.

Using fig.~\ref{fig:PV1}, we can understand why the EFT limit on $\Lambda$ is too large for light mediators. In this figure, we have fixed $g=g_q=g_{\chi}$ and $m_{\rm{DM}}=10$~GeV. The blue dotted and red solid lines correspond to $m_{\rm{med}}=100$~GeV and~1000~GeV, which lie in Region~III and Region~II respectively (c.f.\ fig.~\ref{fig:ratio}). For comparison, the double-dotted black line shows the EFT result for $\Lambda=1000$~GeV. For all three lines, the value of $\Lambda$ is the same.
The left panel demonstrates that the EFT differential cross-section is dramatically different from the cross-section for $m_{\rm{med}}=100$~GeV. The EFT result extends to much larger values of $Q$, the (unobservable) invariant mass of the dark matter pair. This has the effect that the MET distribution for light mediators is much softer, as demonstrated by the blue dotted line in the right panel so that fewer events pass the $\rm{MET}>400$~GeV cut imposed by CMS. This leads to a weaker limit on $\Lambda$. While for $m_{\rm{med}}=1000$~GeV the differential cross-section with respect to $Q$ is also suppressed relative to the EFT cross-section for very large values of $Q$, the MET distribution in both cases is similar.

\section{Comparing CMS and direct detection limits}
\label{sec:DD}

The vector and axial-vector operators constrained by monojet searches also lead to signals at dark matter direct detection experiments. The dark matter in our galaxy is non-relativistic and in this limit, the scattering rates for vector and axial-vector operators lead to `spin-independent' and `spin-dependent' scattering respectively.\footnote{The axial-vector operator also gives a loop-level spin-independent interaction but this is suppressed compared to the spin-dependent interaction~\cite{Haisch:2013uaa}.} Spin-independent scattering is enhanced by coherence effects over the entire nucleus leading to a scattering cross-section proportional to the square of the atomic number $A^2$ (assuming the dark matter couples equally to protons and neutrons). In comparison, spin-dependent scattering is suppressed because the net nuclear spin of isotopes is small and typically, only a fraction of nuclear isotopes carry non-zero spin. Although the usual expectation is that dark matter will first show up through a spin-independent interaction, many exceptions are known where the spin-dependent interaction is larger (see e.g.~\cite{Cohen:2010gj,Freytsis:2010ne,Chalons:2012xf}).

\begin{figure}[t]
\centering
\includegraphics[width=0.48 \columnwidth]{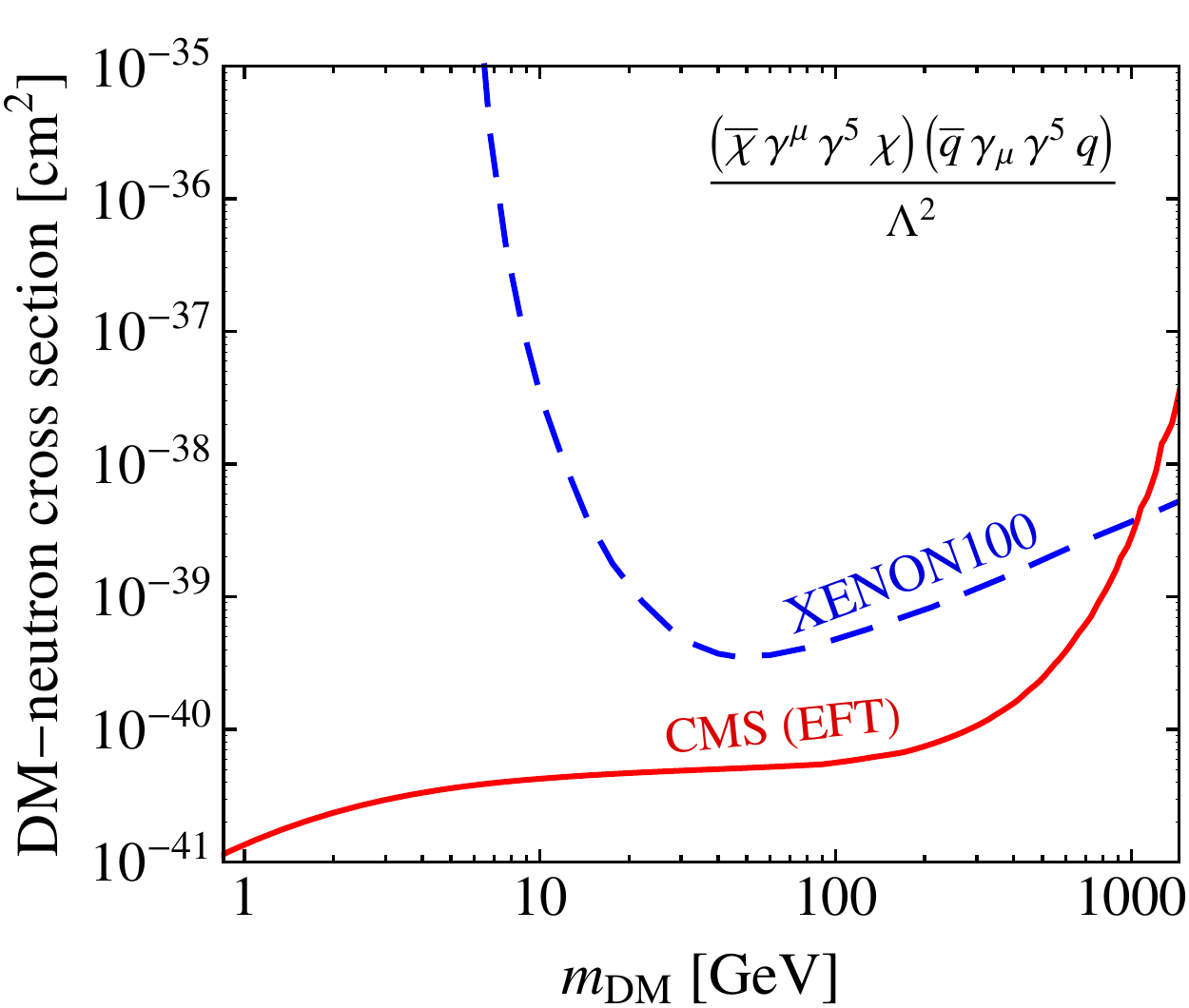}\hspace{3mm}\includegraphics[width=0.48 \columnwidth]{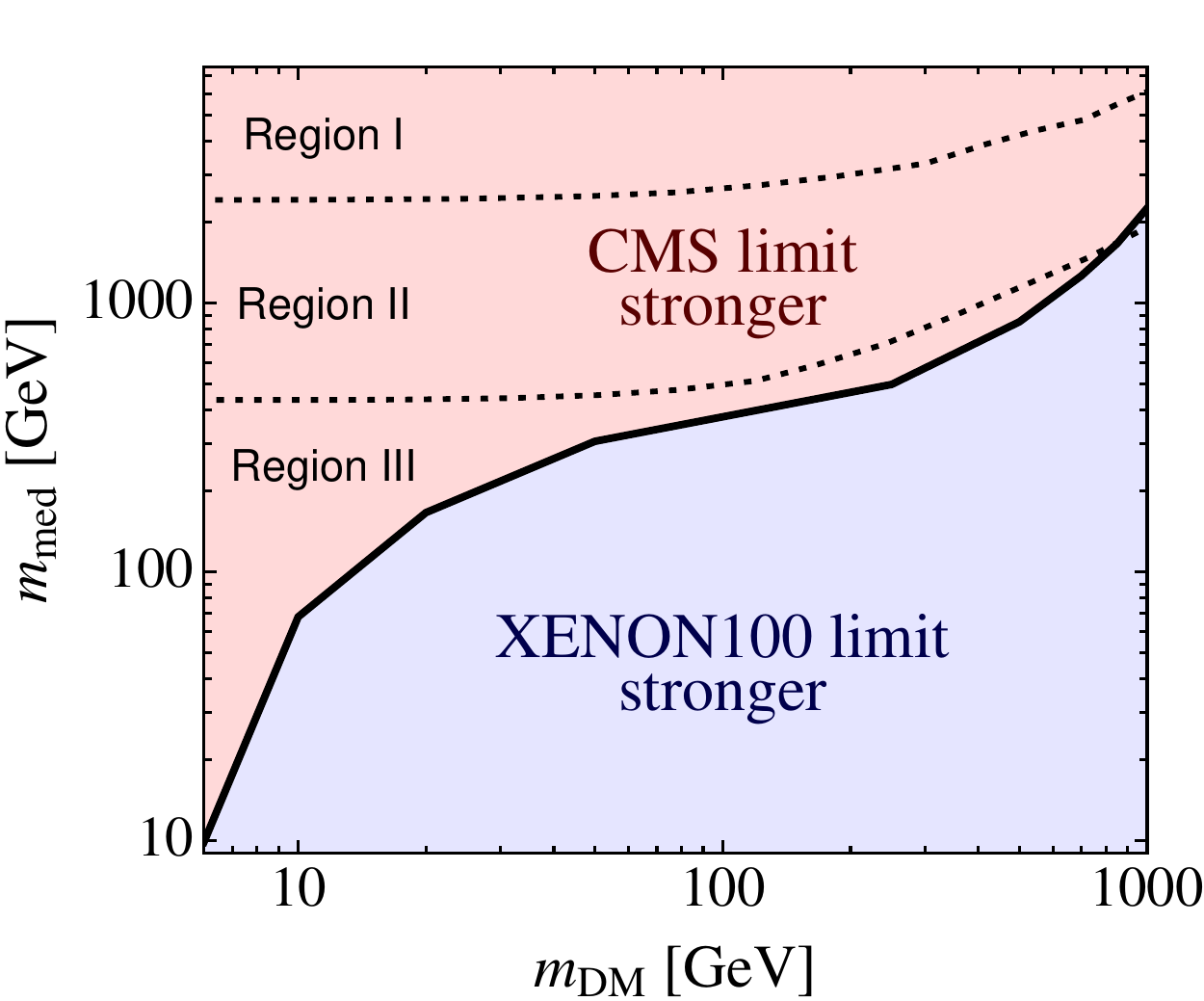}
\includegraphics[width=0.48 \columnwidth]{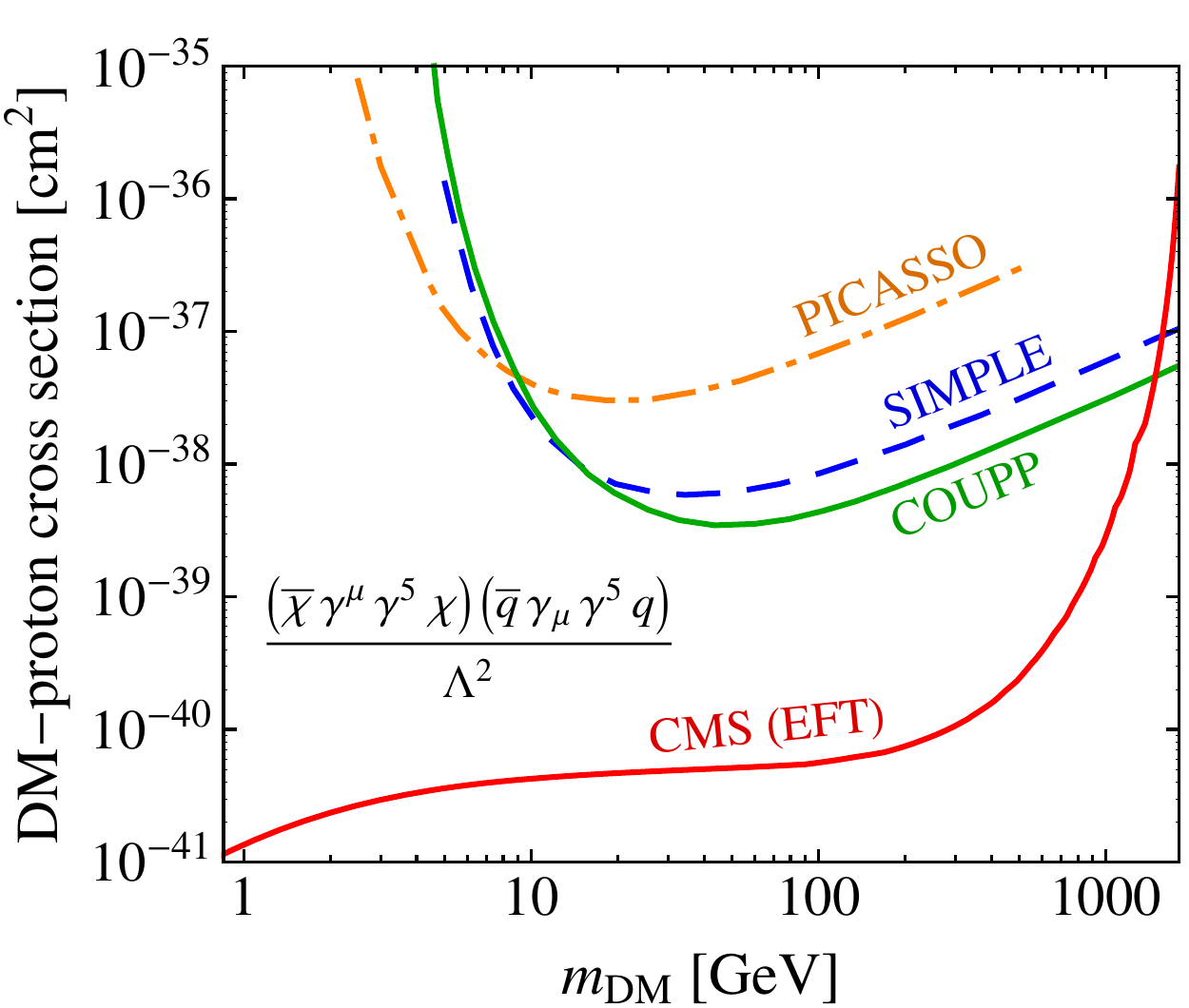}\hspace{3mm}\includegraphics[width=0.48 \columnwidth]{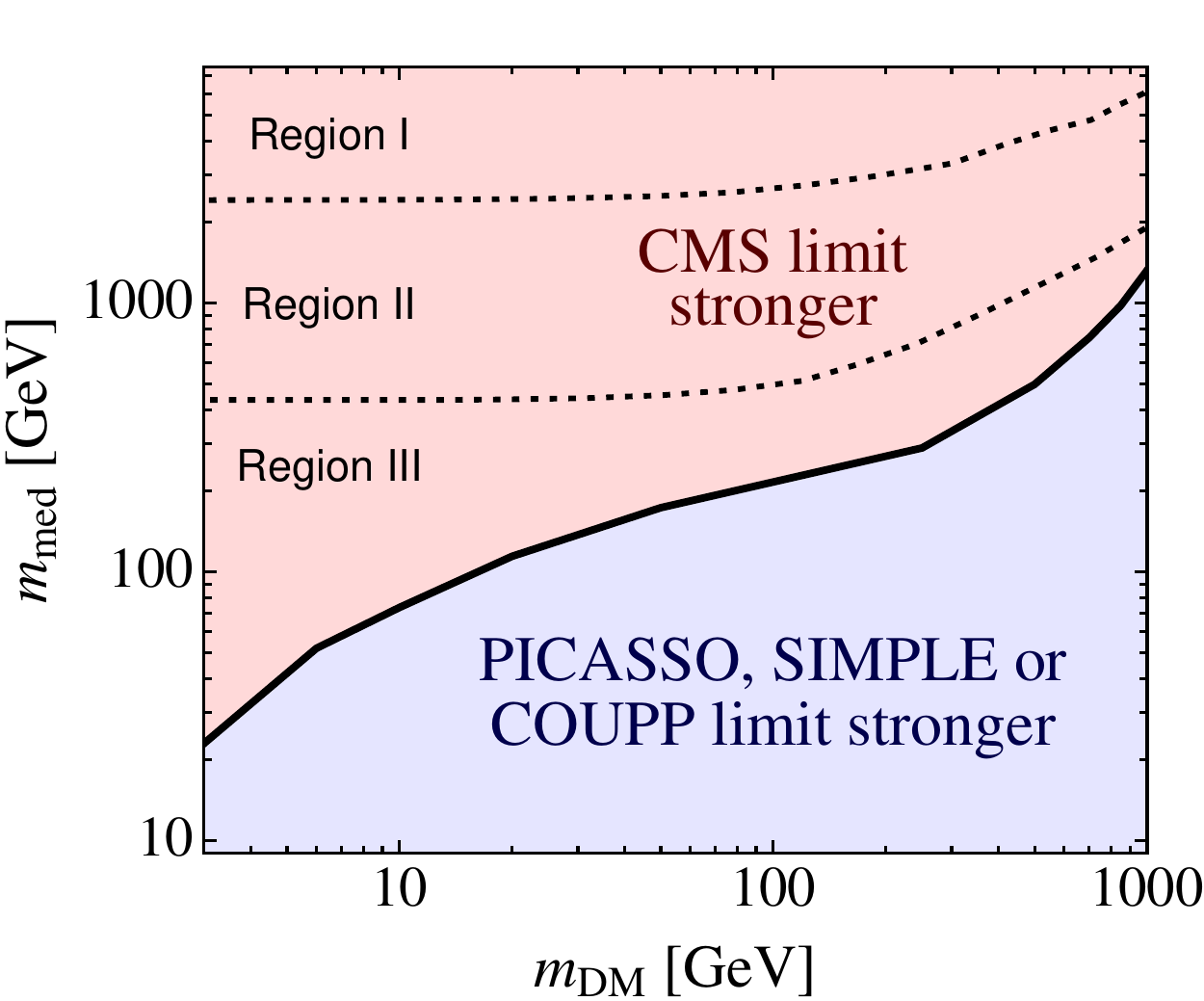}
\caption{
Left panels: The upper (lower) panels show the limits on the spin-dependent dark matter-neutron (proton) scattering cross-section. The solid red line in both panels is the CMS EFT limit. Limits from the XENON100, PICASSO, SIMPLE and COUPP direct detection limits are also shown. When the EFT limit is valid, the CMS EFT limit is stronger than the direct detection limits for $m_{\rm{DM}}\lesssim 1$~TeV. Right panels: The black solid line in the upper (lower) panels indicates the mediator mass $m_{\rm{med}}$ for which the CMS and direct detection dark matter-neutron (proton) scattering cross-section limits are equal. For larger (smaller) $m_{\rm{med}}$, the CMS (direct detection) limit is stronger. The dotted lines distinguish Regions~I,~II and~III. In this range of $m_{\rm{DM}}$, direct detection experiments set a stronger limit in Region~III only.}
\label{fig:SD1}
\end{figure}

The different spin structure between the vector and axial-vector interactions does not lead to a large difference in the relativistic limit. As a result, the LHC sets similar limits on $\Lambda$ for both interactions. From the axial-vector limit on $\Lambda$, the spin-dependent cross-section to scatter off a nucleon $N$ (either a proton or neutron) is given by
\begin{align}
\label{eq:DDxsec1}
\sigma_{\rm{SD}}^N&=\frac{3 \mu_{\rm{red}}^2}{\pi}\frac{a_{N}^2}{\Lambda^4}\\
&\approx5.8\times10^{-41}~\mathrm{cm}^2 \times \left(\frac{\mu_{\rm{red}}}{1~\mathrm{GeV}}\right)^2 \left( \frac{900~\mathrm{GeV}}{\Lambda} \right)^4\;,
\label{eq:DDxsec2}
\end{align}
where $\mu_{\rm{red}}=m_{\rm{DM}} m_N/(m_N+m_{\rm{DM}})$ is the reduced mass of the dark matter-nucleon system and $a_N=\sum_{q=u,d,s}\Delta_q^{N}$ is the coupling of the dark matter to the nucleon spin. We take $\Delta_q^{N}$ from~\cite{Airapetian:2007mh}, leading to $a_p=a_n=0.33$. Equation~\eqref{eq:DDxsec1} is valid when the mediator mass is above 100~MeV (see e.g.~\cite{Chang:2009yt,Frandsen:2013cna})).

The direct detection experiments typically quote their results on the scattering cross-section for neutrons and protons separately. The dashed blue line in the upper left panel of fig.~\ref{fig:SD1} shows the XENON100 90\% CL limit on the cross-section to scatter off a neutron~\cite{Aprile:2013doa}. The solid green, dashed blue and dot-dashed orange lines in the lower left panel of fig.~\ref{fig:SD1} show the 90\% CL limits on the cross-section to scatter off a proton from the COUPP~\cite{Behnke:2012ys}, SIMPLE~\cite{Felizardo:2011uw} and PICASSO~\cite{Archambault:2012pm} experiments. Also shown by the solid red line in both panels is the CMS EFT 90\% CL limit. This was obtained by using the limit on $\Lambda$ in the EFT framework (the solid red line from the left panel of fig.~\ref{fig:CMS1}) in eq.~\eqref{eq:DDxsec2}. That the CMS limit is stronger than the direct detection limits for $m_{\rm{DM}}\lesssim1$~TeV has received much attention.

However, in the previous section we saw that the EFT limit on $\Lambda$ only applies to rather baroque theories of dark matter with a very heavy (and very broad) mediator - we called this Region I. In Region II, the limit on $\Lambda$ is always larger than the EFT limit (see fig.~\ref{fig:ratio}), which implies that the limit on the scattering cross section is stronger than the CMS line in fig.~\ref{fig:SD1}. In Region III, the limit on $\Lambda$ is weaker than the EFT result. Therefore, in this region, the CMS limit on the scattering cross-section will be weaker than the EFT limit and eventually, will be weaker than the direct detection limits in fig.~\ref{fig:SD1}.

In the right panels of fig.~\ref{fig:SD1} we have divided the $m_{\rm{med}}$\,--\,$m_{\rm{DM}}$ plane into regions where the CMS limit is stronger than the direct detection limits (red shaded regions above the thick black line) and {\it vice versa} (blue shaded regions below the thick black line). As expected, it is only in Region~III (where the limit on $\Lambda$ is reduced relative to the EFT limit) that the direct detection experiments set a stronger limit than the collider-based limit. For $m_{\rm{DM}}\lesssim\text{few GeV}$, the CMS limit is always stronger because the direct detection experiments do not place a limit here. Conversely, for $m_{\rm{DM}}\gtrsim1$~TeV, the direct detection limits are always stronger because CMS loses sensitivity here. We have checked this result for $\Gamma=m_{\rm{med}}/8\pi$ and~$m_{\rm{med}}/3$: the difference is so small that it cannot be resolved in fig.~\ref{fig:SD1}.

\subsection{Recommendation for interpretation of monojet analyses in the context of dark matter searches}
\label{sec:recommend}
The canonical interpretation of dark matter searches in the plane of SI/SD direct detection cross-section versus $m_{\rm{DM}}$ is sufficient to fully characterise the results of direct detection experiments. However, more care needs to be taken for the corresponding interpretation of collider based monojet searches. As we have demonstrated in this article, a comprehensive characterisation of monojet searches is governed by four key parameters: the dark matter candidate mass $m_{\rm{DM}}$, the mediator mass $m_{\rm{med}}$ and its width $\Gamma$, as well as the scale $\Lambda$. Only when the dependence of the search result is known for all four of these parameters can an accurate and reliable interpretation in the context of dark matter searches be provided.

We therefore propose that collider experiments provide upper limits on $\Lambda$ from their monojet analyses for each point in the $m_{\rm{DM}}$\,--\,$m_{\rm{med}}$ plane. Since the dependence of $\Lambda$ on $\Gamma$ is confined to the resonance region (see the left panel of fig.~\ref{fig:ratio}), it seems sufficient to repeat this exercise for a few representative values of it (e.g.\ $\Gamma/m_{\rm{med}}=1/8 \pi$, $1/10$ and $1/3$). This strategy is similar to the one already utilised by the experiments in order to characterise searches for supersymmetry in simplified models. There, experimental limits are usually provided in the plane defined by the mass of a sparticle (e.g.\ gluino) and the mass of the lightest supersymmetric particle. In each point of this plane, upper limits on the production cross-section of the sparticle in question are provided. This defines the complete set of information required to characterise this simplified model for searches for direct supersymmetric particle production.

We believe that providing results from monojet searches in this format will enable a complete and reliable interpretation of these analyses in the context of direct dark matter searches. This will also enable comparisons with direct dark matter searches on an equal footing.

\section{Conclusions}
\label{sec:conc}

Monojets are a generic collider-signature in many models of particle dark matter and constitute an important part of the search program for Beyond-the-Standard-Model physics at the LHC. So far, results of these searches have been interpreted in the context of an effective field theory (EFT) framework, which assumes a contact interaction between quarks and dark matter. In this article, we use a simplified model of an (axial)-vector mediator to establish the region of parameter space where the EFT limits on the axial-vector contact operator are valid (the results are similar for the vector contact operator). Throughout, we use the latest monojet analysis from the CMS collaboration~\cite{Chatrchyan:2012me,CMS-PAS-EXO-12-048} as representative of the whole class of monojet searches. In fig.~\ref{fig:CMS1}, we demonstrated that we can reproduce their limits on $\Lambda=m_{\rm{med}}/\sqrt{g_q\, g_{\chi}}$ to better than 15\% accuracy.

When the momentum transfer and mediator mass are similar, the approximation of a contact interaction breaks down and effects from the simplified model must be taken into account (see fig.~\ref{fig:SMS}). Our simplified model for the (axial-)vector mediator is described in section~\ref{sec:valid}. We found that the parameter space divides into three regions (see fig.~\ref{fig:ratio}). In Region~I, the mediator is very heavy, $m_{\rm{med}}>2.5$~TeV, and the limits on the contact interaction scale $\Lambda$ in the EFT apply. However, we showed in fig.~\ref{fig:couplings} that the EFT limit only applies to theories with large couplings $g_q\, g_{\chi}$. While these theories are perturbative for $m_{\rm{DM}}<800$~GeV, the mediator width is larger than the mass $\Gamma>m_{\rm{med}}$ for all values of $m_{\rm{DM}}$. Thus, a particle-interpretation of the mediator is doubtful. Furthermore, assuming the thermal freeze-out mechanism with this mediator, we find a limited mass range $170\lesssim m_{\rm{DM}}\lesssim520$~GeV where the dark matter relic density agrees with the observed value. In Region~II, production of the dark matter pair is resonantly enhanced and the EFT limit on $\Lambda$ underestimates the true value. In contrast, the EFT limit on $\Lambda$ overestimates the true value in Region~III. This is because the MET distribution is softer for light mediators so that fewer events pass the CMS MET cut (see fig.~\ref{fig:PV1}).

The axial-vector CMS limit on $\Lambda$ can be mapped onto the spin-dependent scattering cross-section constrained by direct detection experiments (see fig.~\ref{fig:SD1}). While the cross-section versus $m_{\rm{DM}}$ plane is sufficient to interpret direct dark matter searches, there are two additional parameters for monojet searches: the mediator mass $m_{\rm{med}}$ and width $\Gamma$. Our analysis of the relative sensitivities of monojet and dark matter direct detection searches reveals that both searches dominate in different regions of the $m_{\rm{DM}}$\,--\,$m_{\rm{med}}$ plane (right panel of fig.~\ref{fig:SD1}). Direct detection experiments generally set stronger limits than monojet searches for low mediator masses, where the dependence of the collider limit on $\Gamma$ is small. The monojet searches generally do better at larger mediator masses, where the dependence on $\Gamma$ is more important. Comparing only the EFT limit with direct searches is misleading and can lead to incorrect conclusions about the relative sensitivity of the two search approaches. In fact, our result clearly demonstrates the complementarity of collider-based and direct-detection dark matter searches. Only when both are combined can a comprehensive coverage of the relevant parameter space for dark matter models be achieved. Both search approaches therefore play a critical role in our quest for understanding the nature of dark matter. This conclusion is by no means obvious when based only on the results of the EFT interpretation, in which monojet searches seem to outperform direct detection experiments over a large region in dark matter candidate mass.

\acknowledgments

OB thanks Sarah Malik for useful discussions regarding the CMS monojet analysis. CM and MJD thank Celine B\oe hm, Felix Kahlhoefer, Michael Spannowsky and James Unwin for discussions and Ciaran Williams for help with MCFM. The work of OB is supported in part by the London Centre for Terauniverse Studies (LCTS), using funding from the European Research Council via the Advanced Investigator Grant 267352.

\appendix

\section{`Rules of thumb'}
\label{sec:thumb}

\begin{table}[t]
\centering
\begin{tabular}{@{}l l l l l l l @{}}\toprule 
Quantity &A-V/V & $m_{\rm{DM}}$~[GeV] & $\Gamma/m_{\rm{med}}$& Rule of thumb & Actual & Difference \\ \midrule
$\Lambda_{\rm{peak}}^{\rm{II}}~[\rm{GeV}]$ & A-V & $250$ & $1/8 \pi$ & 1720 &1880 &9\% \\
& A-V& $250$& $1/3$ & 1010 &1050 &4\% \\ 
 & V& $50$& $1/8\pi$ & 2020 &2200 &8\% \\
& V& $50$ & $1/3$ & 1180 &1200 &2\% \\
& V& $500$& $1/10$ & 1280 &1600 &20\% \\ \midrule
Quantity &A-V/V & $m_{\rm{DM}}$~[GeV] & $m_{\rm{med}}$~[GeV]& Rule of thumb & Actual & Difference \\ \midrule
$\Lambda^{\rm{III}}~[\rm{GeV}]$ & A-V & $250$ & $50$ & 70 &50 &40\% \\
 & A-V & $250$ & $100$ & 120 &80 &50\% \\
  & A-V & $250$ & $250$ & 300 &200 &50\% \\
  & V& $50$& $50$ & 110 &110 &0\% \\ 
    & V& $50$& $100$ & 220 &250-370 &12-41\% \\ 
& V& $500$& $50$ & 30 &30 &0\% \\ 
& V& $500$& $100$ & 70 &70 &0\% \\ 
& V& $500$& $500$ & 330 &330 &0\% \\  \bottomrule

\end{tabular} 
\caption{In the upper and lower segments, the difference between the `rule of thumb' and the actual limit on $\Lambda$ for $\Lambda_{\rm{peak}}^{\rm{II}}$ and $\Lambda^{\rm{III}}$ is quantified. The `rules of thumb' give a limit which is accurate to better than $50\%$ accuracy. The shorthand A-V and V refer to axial-vector and vector interactions respectively. The case where a range for $\Lambda$ is quoted reflects the slight dependence of $\Lambda$ on the mediator width.
\label{tab:thumb}}
\end{table}

In section~\ref{sec:valid}, we briefly referred to a number of `rules of thumb'. In the absence of a proper numerical analysis of the monojet search, these can be used to give a quick estimate of the limit on $\Lambda$ for any mediator mass $m_{\rm{med}}$ and dark matter mass $m_{\rm{DM}}$ from knowledge of the EFT limit on $\Lambda$. Here, we collect these rules together and quantify their accuracy.

Firstly, we give an estimate of the boundaries between Regions~I, II and~III . As noted in~\cite{Fox:2012ru}, the resonant enhancement of the signal occurs when $m_{\rm{med}}^2 \gtrsim 4 m_{\rm{DM}}^2 + \slashed{E}^2_T$, where for the CMS search,
$\slashed{E}_{\rm{T}}=400$~GeV. An estimate of the boundary between Regions~II and~III is therefore
\begin{equation}
m_{\rm{med}}^{\rm{II-III}}\approx \sqrt{4 m_{\rm{DM}}^2+(400~\text{GeV})^2}\;.
\end{equation}
Comparing this with fig.~\ref{fig:ratio}, we find that this relationship is accurate to better than $45\%$ over the range $1\lesssim m_{\rm{DM}}\lesssim1000$~GeV.
The position of the boundary between Regions~I and~II is given by requiring that the terms $\mathcal{O}(Q^2/m_{\rm{med}}^2)$ are small (see eq.~\eqref{eq:prop1}). With $Q^2\approx4 m_{\rm{DM}}^2 + \slashed{E}^2_T$, we find that a good estimate of the boundary between Regions~I and~II, which is accurate to better than $10\%$ over the range $1\lesssim m_{\rm{DM}}\lesssim1000$~GeV, is
\begin{equation}
m_{\rm{med}}^{\rm{I-II}}\approx6 \,\sqrt{4 m_{\rm{DM}}^2+(400~\text{GeV})^2}\;.
\end{equation}

Next, we turn to estimating the limit on $\Lambda$ in the three regions. The limit $\Lambda_{\rm{EFT}}^{\rm{I}}$ in Region~I is the limit given by the experimental collaborations (this is the limit in left panel of fig.~\ref{fig:CMS1} for the axial-vector operator). As discussed in section~\ref{sec:valid}, $\Lambda_{\rm{EFT}}^{\rm{I}}$ is independent of $m_{\rm{med}}$ for $m_{\rm{med}}>m_{\rm{med}}^{\rm{I-II}}$. In Region~II, we find that a good estimate of the limit at the peak of the resonance is 
\begin{equation}
\Lambda_{\rm{peak}}^{\rm{II}}\approx\left(\frac{\Gamma}{m_{\rm{med}}}\right)^{-\frac{1}{4}} \Lambda_{\rm{EFT}}^{\rm{I}}
\label{eq:app:II}
\end{equation}
so that, in Region~II, the limit is bounded between $\Lambda_{\rm{EFT}}^{\rm{I}}\leq\Lambda\leq\Lambda_{\rm{peak}}^{\rm{II}}$. We defined the boundary between Regions~II and~III to be mediator mass for which $\Lambda=\Lambda_{\rm{EFT}}^{\rm{I}}$. In section~\ref{sec:lightmed} we observed that the limit on $\sqrt{g_q\, g_{\chi}}$ is approximately constant for $m_{\rm{med}}<m_{\rm{med}}^{\rm{I-II}}$. Therefore, we find that
\begin{equation}
\Lambda^{\rm{III}}\approx m_{\rm{med}}\,\frac{\Lambda_{\rm{EFT}}^{\rm{I}} }{m_{\rm{med}}^{\rm{II-III}}}
\label{eq:app:III}
\end{equation}
in Region~III.

In the upper and lower segments of table~\ref{tab:thumb}, we have compared the difference between the `rule of thumb' and the actual limit on $\Lambda$ for $\Lambda_{\rm{peak}}^{\rm{II}}$ and $\Lambda^{\rm{III}}$ respectively. The shorthand A-V and V refer to axial-vector and vector interactions respectively. We find that for these cases, the `rules of thumb' give a limit which is accurate to $50\%$ accuracy or better. The case where a range for $\Lambda$ is quoted reflects the slight dependence of $\Lambda$ on the mediator width.

These `rules of thumb' for estimating $\Lambda$ can also be used to estimate the CMS constraint on the direct detection scattering cross-section for lighter mediator masses. These cross-sections should be accurate within a factor of two or three (similar to the uncertainty from the local dark matter density~\cite{McCabe:2010zh}). In Region~II, the smallest cross-section (corresponding to the peak value of $\Lambda$) is related to the EFT cross-section $\sigma_{\rm{EFT}}^{\rm{I}}$ by
\begin{equation}
\sigma_{\rm{peak}}^{\rm{II}}\approx\left( \frac{\Gamma}{m_{\rm{med}}}\right)\sigma_{\rm{EFT}}^{\rm{I}}\;.
\end{equation}
Similiary, in Region~III, an estimate for the cross-section is
\begin{equation}
\sigma^{\rm{III}}\approx \left( \frac{m_{\rm{med}}^{\rm{II-III}}}{m_{\rm{med}}}\right)^4 \sigma_{\rm{EFT}}^{\rm{I}}\,.
\end{equation}
Both of these relations are obtained from eqs.~\eqref{eq:app:II} and~\eqref{eq:app:III}, using $\sigma\propto\Lambda^{-4}$.

\section{Relic abundance calculation}
\label{app:relic}


We assume that dark matter achieves it abundance through the thermal freeze-out mechanism~\cite{Zeldovich1,Zeldovich2,Chiu1}. This is a flexible mechanism allowing dark matter with mass ranging from MeV to multi-TeV~\cite{Boehm:2013jpa,Griest:1989wd}. We calculate the relic abundance using the standard approximations~\cite{Steigman:1979kw,Bernstein:1985th,Scherrer:1985zt}. The key quantity determining the relic abundance is the thermally-averaged annihilation cross-section $\langle \sigma v_{\rm{M\o l}}\rangle$, where $v_{\rm{M\o l}}$ is the M\o ller velocity. It was shown in~\cite{Gondolo:1990dk} that $\langle \sigma v_{\rm{M\o l}}\rangle=\langle \sigma v_{\rm{lab}}\rangle \neq \langle \sigma v_{\rm{cm}}\rangle$, where $v_{\rm{lab}}$ and $v_{\rm{cm}}$ are the relative velocities in the lab frame (where one of the incoming dark matter particles is at rest) and the centre of mass frame respectively. Expanding $\sigma v_{\rm{lab}}=a+b v_{\rm{lab}}^2+\mathcal{O}(v_{\rm{lab}}^4)$, the thermal average is $\langle \sigma v_{\rm{lab}} \rangle=a+6 b x^{-1}$, where $x=m/T$. The dark matter relic density (consisting of equal densities $\chi$ and $\bar{\chi}$) is then
\begin{equation}
\Omega_{\chi\bar{\chi}} h^2=2\times8.77\times 10^{-11}~\mathrm{GeV}^{-2}\frac{x_{\rm{f}}}{g_{\star s}^{1/2}(a+3 b/x_{\rm{f}})}\;.
\end{equation}
Here, $x_{\rm{f}}=m_{\rm{DM}}/T_{\rm{f}}$ is the solution of
\begin{equation}
e^{x_{\rm{f}}}=c(c+2) \sqrt{\frac{45}{8}} \frac{g\, m_{\rm{DM}}m_{\rm{Pl}}}{2 \pi^3 g_{\star}^{1/2}}\frac{a+6b/x_{\rm{f}}}{x_{\rm{f}}^{1/2}}\;,
\end{equation}
where $m_{\rm{Pl}}$ is the Planck mass, $g$, $g_{\star}$ and $g_{\star s}$ are the usual degrees of freedom (see e.g.~\cite{KolbTurner}) and $c$ is a matching constant, which we set to $\tfrac{1}{2}$. For the observed value of $\Omega_{\chi\bar{\chi}} h^2$, we take $\Omega_{\chi\bar{\chi}} h^2=0.119$~\cite{Ade:2013zuv}.

For the axial-vector operator defined in eq.~\eqref{eq:LeffEFT1axvec}, the annihilation cross-section is
\begin{equation}
\sigma v_{\rm{lab}}=\sum_q \frac{3}{2 \pi \Lambda^4}\sqrt{1-\frac{m_q^2}{m_{\rm{DM}}^2}}\left[m_q^2+\frac{v_{\rm{lab}}^2}{24} \frac{23 m_q^4-28 m_q^2 m_{\rm{DM}}^2+8 m_{\rm{DM}}^4}{m_{\rm{DM}}^2-m_q^2} \right]
\label{sigmavEFT}
\end{equation}
where $m_q$ and $m_{\rm{DM}}$ is the quark mass and dark matter mass respectively and the sum extends over all quarks where $m_{\rm{DM}}>m_q$. This result agrees with the results in~\cite{Srednicki:1988ce, Zheng:2010js}. The results in~\cite{Bai:2010hh,Fox:2011fx, Fox:2011pm} and~\cite{MarchRussell:2012hi,Haisch:2013uaa} differ because their expansion is $\sigma v_{\rm{cm}}$ (our result for $\sigma v_{\rm{cm}}$ (not shown) agrees).

In our simplified model of the axial-vector with the interaction Lagrangian
\begin{equation}
\mathcal{L}_{\rm{int}}=-g_{\chi}Z^{'}_{\mu}\bar{\chi}\gamma^{\mu}\gamma^5\chi-g_{q}Z^{'}_{\mu}\bar{q}\gamma^{\mu}\gamma^5 q\;,
\end{equation}
the annihilation cross-section for the $s$-channel process $\bar{\chi}\chi\rightarrow Z^{'} \rightarrow \bar{q}q$ is
\begin{equation}
\begin{split}
\sigma v_{\rm{lab}}& =\sum_q \frac{3 g_{\chi}^2 g_q^2}{2 \pi m_{\rm{med}}^{4}}\sqrt{1-\frac{m_q^2}{m_{\rm{DM}}^2}}\left[\left(1-\frac{4 m_{\rm{DM}}^2}{m_{\rm{med}}^2}\right)^{2}+\frac{\Gamma^2}{m_{\rm{med}}^2}\right]^{-1}\\
\qquad &\times \left[m_q^2+\frac{v_{\rm{lab}}^2}{24}\left( \frac{23 m_q^4-28 m_q^2 m_{\rm{DM}}^2+8 m_{\rm{DM}}^4}{m_{\rm{DM}}^2-m_q^2} +\frac{48 m_{\rm{DM}}^2 m_q^2 (m_{\rm{med}}^2-4 m_{\rm{DM}}^2)}{(m_{\rm{med}}^2-4 m_{\rm{DM}}^2)^2+m_{\rm{med}}^2 \Gamma^2} \right) \right]\;.
\end{split}
\end{equation}
Again, the sum extends over all quarks where $m_{\rm{DM}}>m_q$. Here $m_{\rm{med}}$ and $\Gamma$ are the mediator mass and mediator width respectively. This result agrees with eq.~\eqref{sigmavEFT} in the limit that $m_{\rm{med}}$ is very large (and when $\Lambda\equiv m_{\rm{med}}/\sqrt{g_{\chi}g_q}$ is identified). We assumed that $\Gamma=m_{\rm{med}}$ in the calculation of the green dot-dashed line in fig.~\ref{fig:couplings}.

\section{Simplified models for scalar mediators}
\label{app:scalar}

\begin{figure}[h]
\centering
\includegraphics[width=0.45\columnwidth]{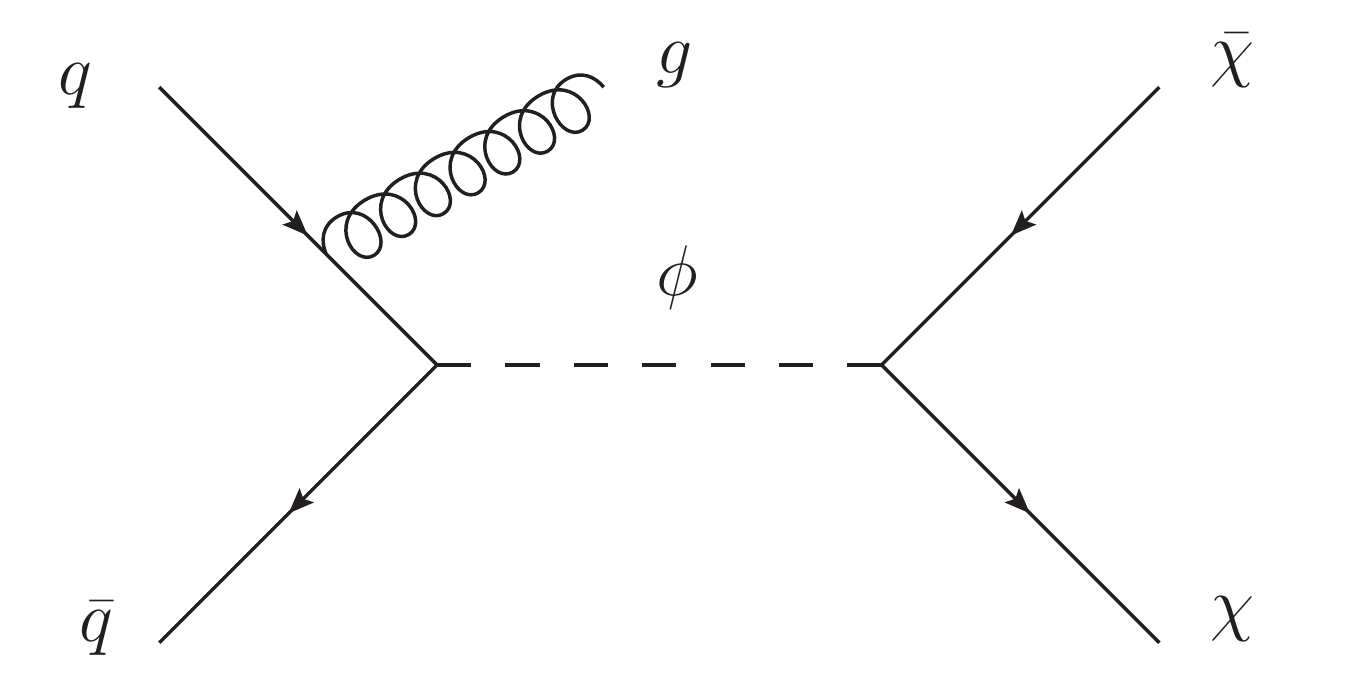}
\includegraphics[width=0.53\columnwidth]{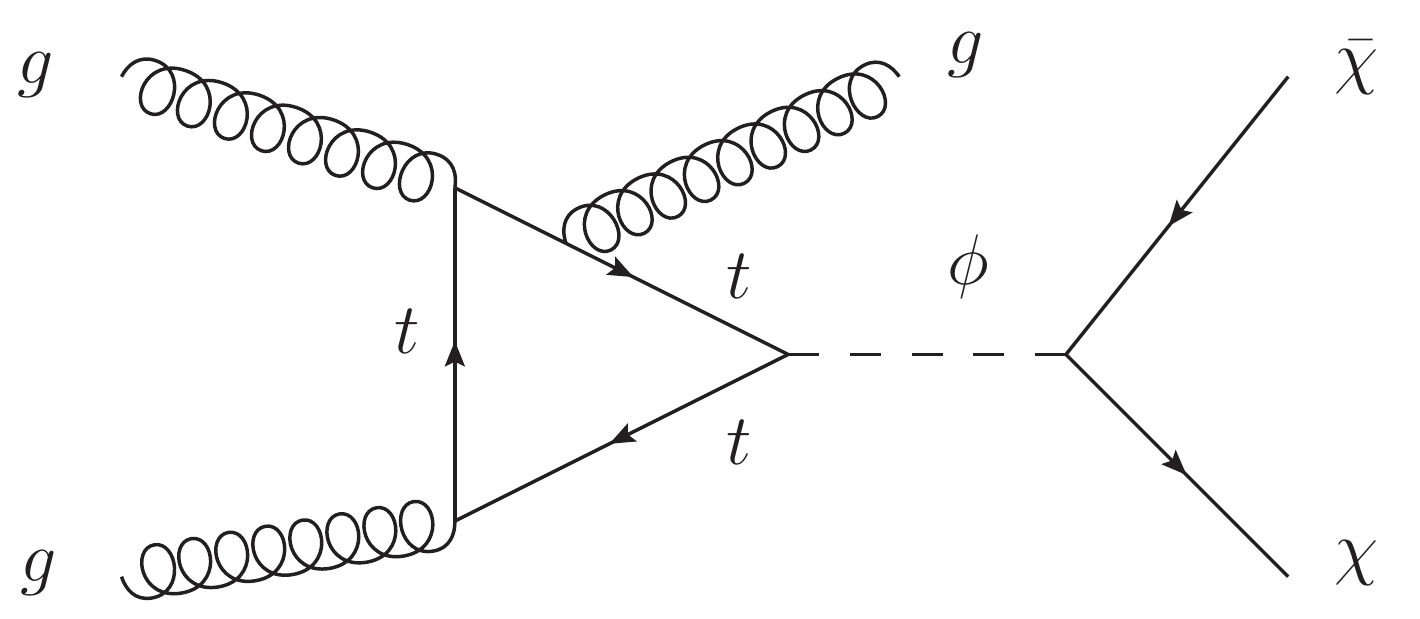}
\caption{Monojet production for scalar mediators. Left panel: A scalar mediator is produced at tree-level from valence quarks. This is suppressed because the valence quark Yukawa couplings are small. Right panel: The dominant production comes from the one-loop gluon fusion contribution with a top-quark in the loop.}
\label{fig:SMS2}
\end{figure}

\noindent In this appendix, we briefly discuss simplified models for scalar mediators $\phi$ (scalar mediators have also been discussed in~\cite{Goodman:2011jq,Fox:2012ee,Haisch:2012kf}). We assume that the dark matter is a Dirac fermion $\chi$ and that neither $\chi$ nor $\phi$ is charged under the Standard Model gauge groups. The relevant interaction terms in the Lagrangian (ignoring the kinetic and mass terms) are
\begin{equation}
\mathcal{L}=- \sum_q  \lambda_q \phi \bar{q}(g_{\mathrm{S} q} + g_{\mathrm{PS} q}\gamma^5   )q    -    \phi \bar{\chi} (g_{\mathrm{S} \chi} +  g_{\mathrm{PS} \chi} \gamma^5   )     \chi \;.
\label{eq:scalar_lagrangian}
\end{equation}
The sum is over all quarks, including the top, and $\lambda_q$ is the usual Yukawa coupling. We mention two reasons why the assumption that the mediator couples to the quarks with a coupling proportional to the Yukawa couplings is reasonable. Firstly, the couplings of light scalars often arise via mixing with the Standard Model Higgs. Secondly, flavour-changing-neutral-currents (FCNCs) are naturally suppressed~\cite{Kamenik:2011nb,Cheung:2010zf}.

As emphasised in~\cite{Haisch:2012kf}, the dominant contribution to the cross-section comes from the gluon-fusion loop shown in fig.~\ref{fig:SMS}. This is because the top-Yukawa is much larger than the other Yukawa couplings. Considering only the valence quark contribution gives a limit on $\Lambda$ that is too weak by more than a factor of two~\cite{Haisch:2012kf}. Integrating out the top-loop to obtain an effective operator such as $\phi\, G^{a}_{\mu\nu}G^{a \,\mu\nu}$, where $G^{a}_{\mu\nu}$ is the usual QCD field strength tensor, is a bad approximation because the large missing $E_{\rm{T}}$ and $p_{\rm{T}}$ cuts are at scales larger than the top mass. This leads to a cross-section which is wrong by up to a factor of 40 in some regions of parameter space~\cite{Haisch:2012kf}.

\bibliography{ref}
\bibliographystyle{JHEP}

\end{document}